\documentclass[12pt]{article}
\usepackage{graphicx}
\usepackage{amsmath}
\usepackage{amssymb}
\usepackage{slashed}
\usepackage{bigstrut}
\usepackage{stmaryrd}
\usepackage{mathtools}
\usepackage{multirow}
\newcommand{\beq}{\begin{equation}}
\newcommand{\eeq}{\end{equation}}
\def\bsp#1\esp{\begin{split}#1\end{split}}
\def\bal#1\eal{\begin{align}#1\end{align}}
\newcommand{\beeq}{\begin{eqnarray}}
\newcommand{\eeeq}{\end{eqnarray}}

\newcommand{\helacnlo}{\texttt{HELAC-NLO}}

\newcommand{\amcatnlo}{\texttt{aMC@NLO}}

\newcommand{\pythiaver}{\texttt{PYTHIA-6.4.25}}

\newcommand{\madgraphfive}{\texttt{MADGRAPH5}}

\newcommand{\powheg}{POWHEG}
\newcommand{\powhegbox}{\texttt{POWHEG-Box}}
\newcommand{\powhel}{\texttt{PowHel}}

\newcommand{\fastjet}{\texttt{FastJet}}
\newcommand{\heptoptagger}{\texttt{HEPTopTagger}}
\newcommand{\tq}{\mathrm{q}}

\newcommand{\gev}{\ensuremath{\,\mathrm{GeV}}}

\newcommand{\ud}{\mathrm{d}}

\newcommand{\bB}	{b$\bar{{\rm b}}$}
\newcommand{\tT}	{t$\bar{{\rm t}}$}
\newcommand{\ttH}	{t$\bar{{\rm t}}$\,$H$}
\newcommand{\muf}   {\ensuremath{\mu_{\rm F}}}
\newcommand{\mur}   {\ensuremath{\mu_{\rm R}}}

\newcommand{\pT}{\ensuremath{p_{\perp}}}

\newcommand{\pTj}{\ensuremath{p_{\perp,j}}}

\newcommand{\pTgamma}[1]{\ensuremath{p_{\perp,\,\gamma_{#1}}}}
\newcommand{\pTlepton}[1]{\ensuremath{p_{\perp,\,\ell^{#1}}}}
\newcommand{\pTsupp}{\ensuremath{p_{\perp,{\rm supp}}}}
\newcommand{\pTsuppR}{\ensuremath{p_{\perp,{\rm supp}}^{r}}}
\newcommand{\msupp}{\ensuremath{m_{\rm supp}}}
\newcommand{\maa}{\ensuremath{m_{\gamma\gamma}}}

\newcommand{\kT}{\ensuremath{k_{\perp}}}

\newcommand{\mt}{\ensuremath{m_{\rm t}}}
\newcommand{\mh}{\ensuremath{m_{H}}}

\newcommand{\as}{\ensuremath{\alpha_{\rm s}}}
\newcommand{\aem}{\ensuremath{\alpha_{\rm EM}}}
\newcommand{\rt}{{\rm t}}
\newcommand{\bt}{\ensuremath{\bar{{\rm t}}}}

\newcommand{\uq}	{\ensuremath{\mathrm{u}}}
\newcommand{\uaq}	{\ensuremath{\mathrm{\bar{u}}}}
\newcommand{\dq}	{\ensuremath{\mathrm{d}}}
\newcommand{\daq}	{\ensuremath{\mathrm{\bar{d}}}}
\newcommand{\ttgamma}	{\ensuremath{\rt\,\bt\,\gamma}}
\newcommand{\ttaa}	{\ensuremath{\rt\,\bt\,\gamma\,\gamma}}

\newcommand{\tth}	{\ensuremath{\rt\,\bt\,H}}
\newcommand\fig[1]     {Fig.\,{\ref{#1}}}
\newcommand\figs[2]    {Figs.\,{\ref{#1}} and ~\ref{#2}}
\newcommand\figss[2]   {Figs.\,{\ref{#1}}--\ref{#2}}
\newcommand\sect[1]    {Sect.\,{\ref{#1}}}

\newcommand\tab[1]     {Table~\ref{#1}}

\begin{document}

\begin{titlepage}
\vspace*{-2cm}
\vskip .5in
\begin{center}
{\large\bf
Hadroproduction of t anti-t pair with two isolated photons with PowHel
}\\
\vspace*{1.5cm}
{\large A.~Kardos and Z.~Tr\'ocs\'anyi} \\
\vskip 0.2cm
Institute of Physics and MTA-DE Particle Physics Research Group,\\
  University of Debrecen, H-4010 Debrecen P.O.Box 105, Hungary 
\vskip 1cm
\end{center}

\par \vspace{2mm}
\begin{center} {\large \bf Abstract} \end{center}
\begin{quote}
\pretolerance 10000
We simulate the hadroproduction of a t\bt\ pair in association with two
isolated hard photons at 13\,TeV LHC using the \powhel\ package. We use
the generated events, stored according to the Les-Houches event format,
to make predictions for differential distributions formally at the
next-to-leading order (NLO) accuracy. We present predictions at the
hadron level employing the cone-type isolation of the photons used by
experiments.  We also compare the kinematic distributions to the same
distributions obtained in the t\bt$H$ final state when the Higgs-boson
decays into a photon pair, to which the process discussed here is an
irreducible background.
\end{quote}

\vspace*{\fill}
\begin{flushleft}
July 2014
\end{flushleft}
\end{titlepage}

\section{\label{sec:introduction} Introduction}

The Higgs-boson was discovered by the ATLAS \cite{Aad:2012tfa} and
CMS \cite{Chatrchyan:2012ufa} experiments two years ago, and many of
its properties have been measured since then. The results of these
measurements are in agreement with the predictions of the Standard
Model within the uncertainties of the measurements:
(i) it is a $J^P = 0^+$ particle,
(ii) the branching ratios are as predicted, and 
(iii) it couples to the masses of the heavy gauge bosons
\cite{Aad:2013xqa,Chatrchyan:2013iaa}.  Its mass is not predicted, but
measured consistently by the two experiments:
$m_H/\mathrm{GeV} = 125.5\pm 0.2_{\mathrm{stat}}\pm 0.6_{\mathrm{syst}}$
by ATLAS \cite{Aad:2013wqa} and $125.6\pm 0.4_{\mathrm{stat}}\pm
0.2_{\mathrm{syst}}$ by CMS \cite{Chatrchyan:2013mxa}.

The measurements in order to discover the properties of the Higgs-boson
are not over. It is yet to measure its couplings to the fermions $y_{\rm
f}$ as well as to check if its triple and quartic self couplings are
consistent with its mass. The smallness of these couplings require
large integrated luminosity and such measurements become feasible only
at the designed c.m.~energy of the LHC. In this respect the t-quark is
special among the fermions due to its large mass. As $m_H < m_\tq$, the
Higgs-boson cannot decay into a \tT-pair, therefore, the
Yukawa-coupling $y_\tq$ can only be measured directly from \ttH\ final
states, which is an important goal at the LHC.

Measuring the \ttH\ production cross section is very challenging due
to the small production rates and in general large backgrounds. The
experiments concentrate on studying many decay channels sorted into
three main categories: (i) the hadronic, (ii) the leptonic and the
(iii) di-photon channels. In the hadronic channel the Higgs-boson
is assumed to decay into a \bB\ or into $\tau^+\tau^-$ pair, while one
or both t-quarks decay leptonically (hadrons with single lepton or
dilepton channels). In the leptonic channel the Higgs-boson decays
into charged leptons and missing energy (through heavy vector bosons),
while one or both t-quarks decay again leptonically. Finally, in the
di-photon channel the Higgs-boson decays into a photon pair, while the
t-quarks decay into jets (di-photon with hadrons), or into the
semileptonic channel (di-photon with lepton). Common characteristics of
all these channels is the large background from other SM processes.
Thus a precise measurement needs to be aided by theory through precise
predictions of distributions at the hadron level for the hadroproduction
of a \tT-pair in association with one or two hard object(s) $X$, with
$X=H$ \cite{Garzelli:2011vp}, $Z^0$ \cite{Kardos:2011na,Garzelli:2011is},
$W^\pm$ \cite{Garzelli:2012bn}, an isolated photon \cite{Kardos:2014zba},
jet \cite{Kardos:2011qa}, \bB-pair \cite{Kardos:2013vxa}, or a pair of
jets or isolated photons.  The goal of the present article is to make
high-precision predictions for the last of these.

Isolated hard photons are important experimental tools at the LHC. 
However, in perturbation theory these are rather cumbersome objects
because the photons couple directly to quarks. If the quark that emits
the photon is a light quark, treated massless in perturbative QCD, then
the emission is enhanced at small angles and in fact, becomes singular
for strictly collinear emission. Due to this divergence of the
collinear emission, the usual experimental definition of an isolated
photon, which allows for small hadronic activity even inside the
isolation cone, cannot be implemented directly in a perturbative
computation. One has to take into account the non-perturbative
contribution of photon fragmentation, too \cite{Catani:1998yh}.

Recently, a new method was presented to make predictions for processes
with isolated hard photons in the final state \cite{Kardos:2014zba}. This
method builds on the POWHEG method \cite{Nason:2004rx,Frixione:2007vw}
as implemented in the \powhegbox~\cite{Alioli:2010xd}. The \powhegbox\
program requires input from the user --
the Born phase space and various matrix elements: (i)
the Born squared matrix element (SME), (ii) the spin- and (iii)
color-correlated Born SME, (iv) the SMEs for real and (v) the virtual
corrections. In the \powhel\ framework \cite{Garzelli:2011iu} we obtain
all these ingredients from the \helacnlo\ code \cite{Bevilacqua:2011xh}.
The result of \powhel\ is pre-showered events (with kinematics up to
Born plus first radiation) stored in Les Houches event (LHE) files
\cite{Alwall:2006yp}. These events can be further showered and
hadronized using standard shower Monte Carlo (SMC) programs up to the
hadronic stage, where any experimental cut can be used, so a
realistic analysis becomes feasible.

The essence of the new method of predicting cross sections for
processes with isolated photons is to introduce generation isolation
\cite{Kardos:2014zba} and generation cuts
\cite{Kardos:2011qa,Alioli:2010qp} during event generation that are
sufficiently small so that predictions at the hadron level obtained
with physical cuts and isolation do not depend on them.  If the
generation isolation of the photon is smooth \cite{Frixione:1998jh},
that is infrared safe at all orders of perturbation theory, then the
pre-showered events can effectively be considered as sufficiently
inclusive event sample, which leads to predictions for distributions
obtained with usual experimental isolation. While we neglect the
fragmentation contribution of the photon, `sufficiently inclusive'
means that the physical predictions obtained with typical experimental
isolation employed on these shoewerd LHEs are (i) independent of the
generation isolation and (ii) the fragmentation contribution should be
negligible compared to the scale dependence when the photons are harder
than possible accompanying jets \cite{Kardos:2014zba}.

The new method of making predictions for processes including isolated
photons in the final state at the next-to-leading order (NLO) accuracy
matched with parton shower (PS) have been tested thoroughly for the case
of \ttgamma\ hadroproduction in Ref.~\cite{Kardos:2014zba}. Here we use
the same method for the case of \ttaa\ hadropoduction, an irreducible
background for \tth\ final states in the $H\to \gamma\gamma$ decay
channel. This process was first computed at leading-order accuracy in
Refs.~\cite{Kunszt:1991xk,Ballestrero:1992bk}. Here we make predictions
for this process at NLO accuracy and compare the predictions of \powhel\
to those of \madgraphfive\ using the Frixione isolation. Then we show
that predictions at the hadronic stage are independent of the
generation isolation and cuts and finally make predictions at the LHC
at 13\,TeV collision energy and compare the \ttH($H\to \gamma\gamma$)
signal to the \ttaa\ background.

\section{\label{sec:method} Implementation}

Details of generating LHEs with the \powhegbox\ have been discussed
extensively in the literature. Here we only summarize those features of
our implementation that are specific to the \ttaa\ hadroproduction
process.

The process $p\,p(\bar{p})\to\rt\,\bt\,\gamma\,\gamma$ involves several
subprocesses. Not all of these are independent, but related by
crossing. Hence matrix elements, both at tree-level and at one-loop,
are generated for a subset of subprocesses by \helacnlo\ and all others
are obtained by crossing into relevant channels. In \helacnlo\ the
matrix elements are stored in separate files using an internal
notation, these are commonly dubbed as skeleton files. For this
process, we generated skeleton files for the following subprocesses:
$g\,g\to\gamma\,\gamma\,\rt\,\bt$,
$\uq\,\uaq\to\gamma\,\gamma\,\rt\,\bt$ and 
$\dq\,\daq\to\gamma\,\gamma\,\rt\,\bt$ for the Born and virtual, while 
$g\,g\to\gamma\,\gamma\,\rt\,\bt\,g$,
$\uq\,\uaq\to\gamma\,\gamma\,\rt\,\bt\,g$ and 
$\dq\,\daq\to\gamma\,\gamma\,\rt\,\bt\,g$ for the real-emission
contribution.  The ordering among final state particles is in
compliance with that used by \powhegbox.
We checked the correct evaluation of all the matrix elements by
comparing them in several, randomly picked phase space points to the
original ones in \helacnlo. The self-consistency between real emission
contribution and subtraction terms is established by verifying that the
limiting value of the ratio of the real emission part and the
subtraction terms in all kinematically degenerate configurations
approaches one.

The phase space for \ttaa-production with two massless particles in the
final state at the Born level is quite involved provided by possible
singular photon-radiation from initial state (anti)quarks. Thus we
decided to build the Born phase space in a recursive fashion using the
method of \cite{Campbell:2012am,Kardos:2014dua}.  To construct an
$(n+1)$-particle phase space point from an $n$-particle one we use the
inverse constructions of \cite{Frixione:2007vw} with three random numbers of
unity. A choice can be made between two possible inverse constructions,
initial and final state ones, depending upon the underlying splitting.
Due to the mass of the t-quark, singular photon radiation can only come
from the initial state hence in our phase-space construction we
considered only initial state as a possible source for photons, which we
arranged symmetrically for the two photons.

\section{\label{sec:NLO}Predictions at NLO}

With the internal consistency of matrix elements having checked and a
suitable underlying Born phase space set up, the code is validated to make
predictions at NLO accuracy. With the publicly available independent
programs \madgraphfive/\amcatnlo\ \cite{Alwall:2014hca} we can compare
the predictions of the two independent codes.  For this comparison we
used the following set of parameters: $\sqrt{s} = 8$\,TeV at the LHC,
\texttt{CTEQ6L1} PDF with a 1-loop running \as\ for the LO and
\texttt{CTEQ6M} PDF with a 2-loop running \as\ for the NLO comparison. 
We used equal factorization and renormalization scales set to the mass
of the t-quark, $\mt = 172.5\,\gev$.  We set the fine-structure
constant to $\aem = 1/137$ and kept fixed at that value. We used
five massless-quark flavores throughout the paper.  We employed
the following set of selection cuts: (i) both photons had to be hard,
$\pTgamma{i} > 30\,\gev$, $i\in\{1,2\}$ and (ii) 
central, $|y_{\gamma_i}| < 2.5$, $i\in\{1,2\}$, (iii) a Frixione
isolation \cite{Frixione:1998jh} was applied to the photons with
$\delta_0 < 0.4$, $\epsilon_\gamma = n = 1$. When this isolation
is applied the partonic activity is limited around the photon if the
separation is less than $\delta_0$. The partonic activity in a cone of
$\delta$ around the photon is measured through the total transverse
energy deposited by partons in this cone, this total transverse energy
is limited,
\begin{align}
E_{\bot} = \sum_{i\in{\rm partons}} E_{\bot,\,i}
\Theta\left(\delta - R(p_\gamma,\,p_i)\right)
\le
E_{\bot,\,\gamma}\left(\frac{1 - \cos\delta}{1 - \cos\delta_0}\right)
\label{eqn:smoothisol}
\,,
\end{align}
where $\delta\le\delta_0$. 
Our purpose with this comparison is only to show that the two codes
lead to the same predictions.

\begin{table}
\centering
\begin{tabular}{|c|c|c|}
\hline
\hline
& $\sigma^{\rm LO}$ [ab] & $\sigma^{\rm NLO}$ [ab] \bigstrut\\
\hline
\hline
\amcatnlo & $  863 \pm 1$ & $ 1063 \pm 2$ \bigstrut\\
\hline
\powhel   & $  864 \pm 1$ & $ 1060 \pm 7$ \bigstrut\\
\hline
\hline
\end{tabular}
\caption{{\label{tbl:compMG5xs}} Cross sections at LO and NLO accuracy
obtained with \madgraphfive/\amcatnlo\ and \powhel. We refer to the main text for 
the details of the computations.}
\end{table}

\begin{figure}[t]
\includegraphics[width=0.50\textwidth]{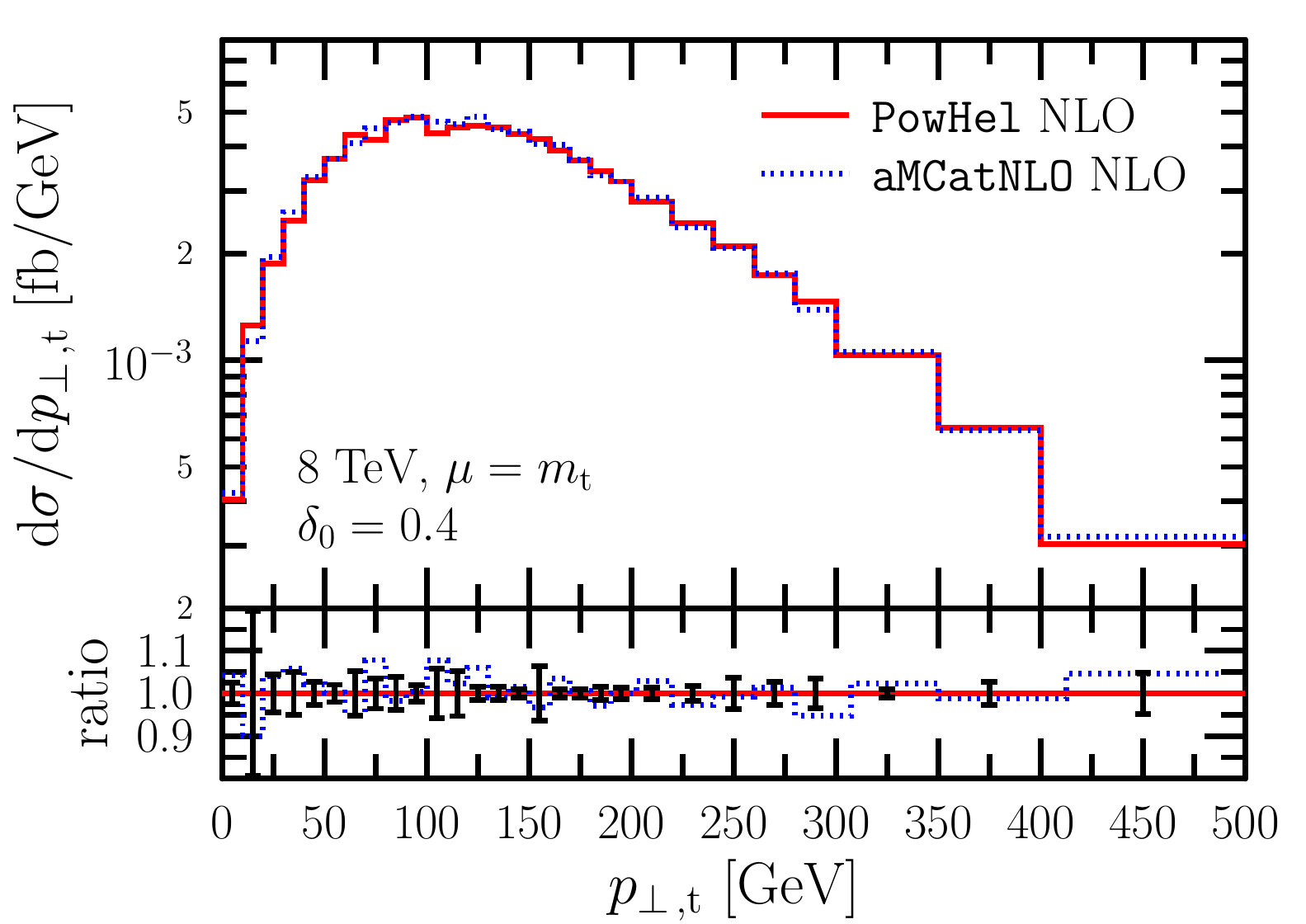}
\includegraphics[width=0.50\textwidth]{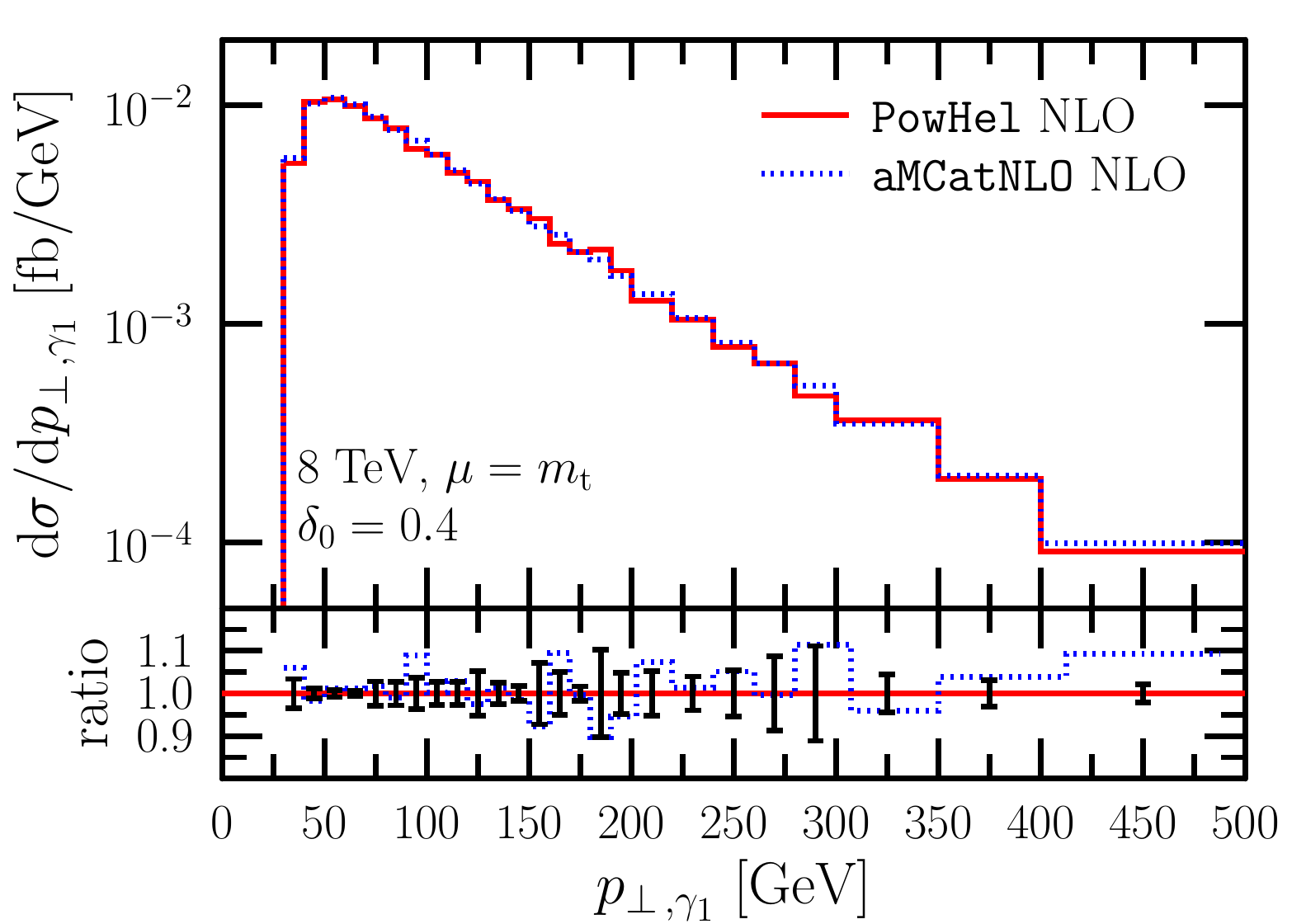}
\caption{\label{fig:NLOcomp-pt-tq-a1} Comparison of predictions at NLO to
those of \madgraphfive/\amcatnlo\ \cite{Alwall:2014hca} for transverse
momentum of the top quark and the hardest photon. The lower panel shows
the ratio of the predictions. For details on the selection cuts we
refer to the text.}
\end{figure}

\begin{figure}[t]
\includegraphics[width=0.50\textwidth]{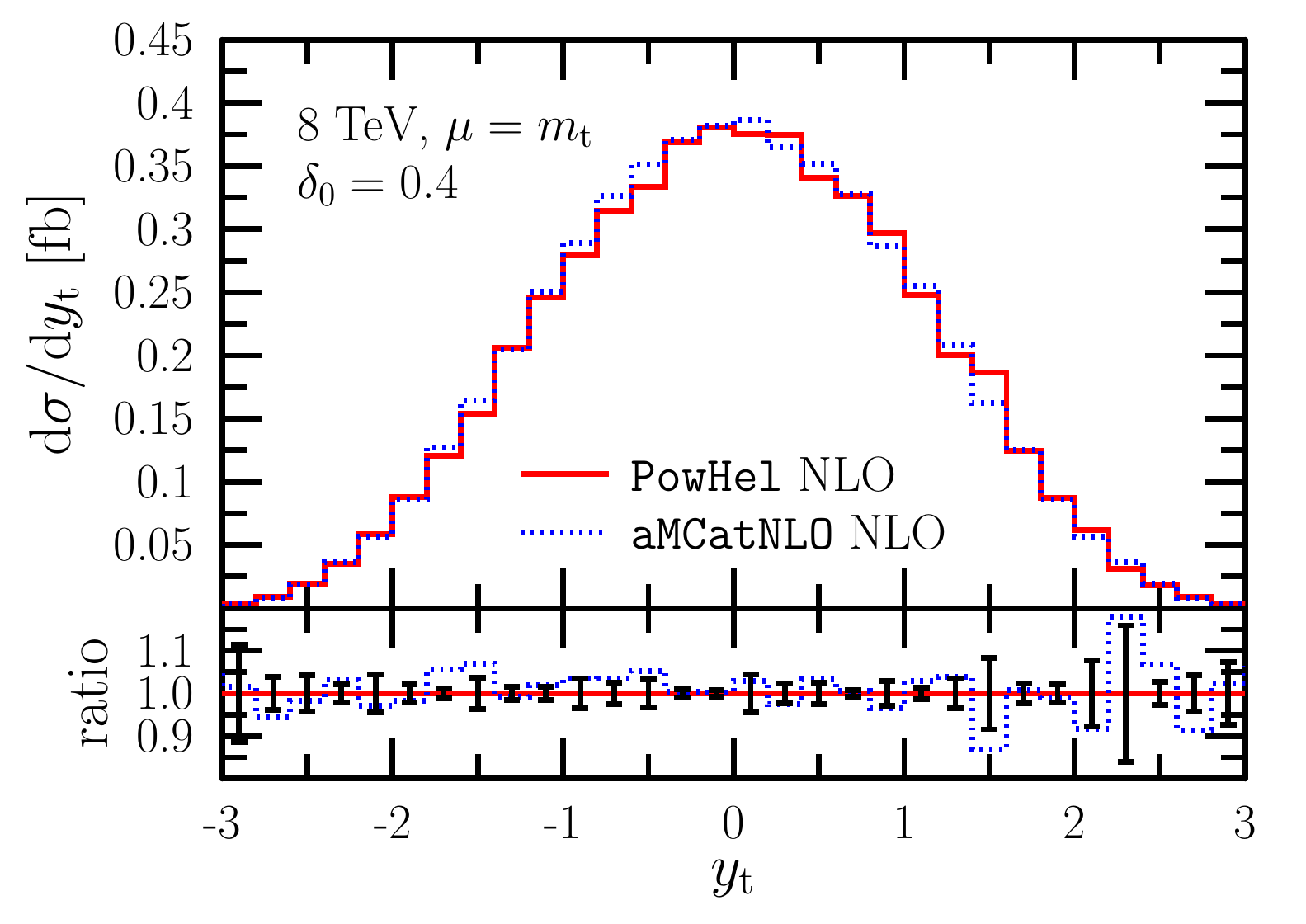}
\includegraphics[width=0.50\textwidth]{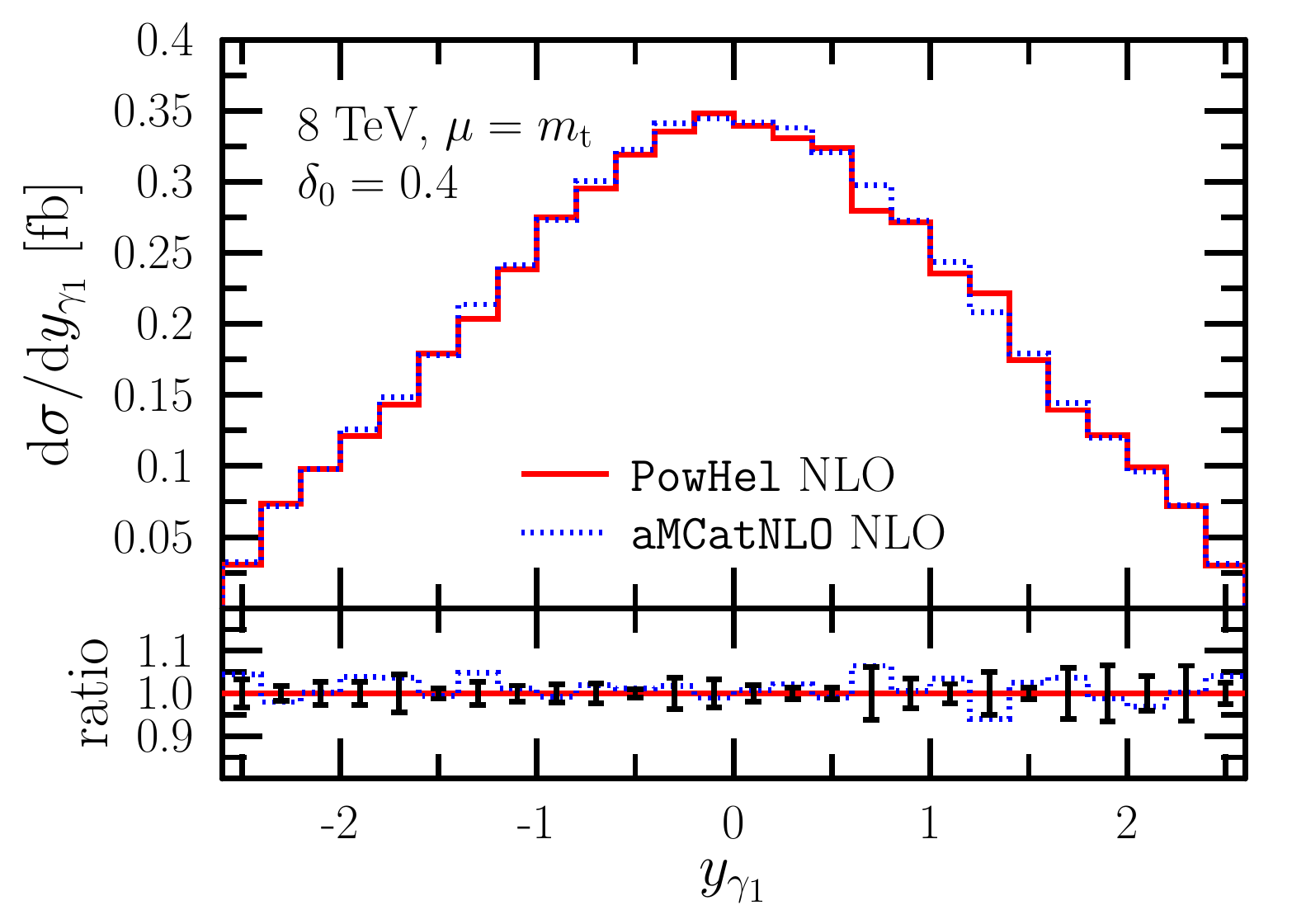}
\caption{\label{fig:NLOcomp-y-tq-a1} The same as \fig{fig:NLOcomp-pt-tq-a1}
but for the top quark and hardest photon rapidities.}
\end{figure}

The cross section corresponding to our selection cuts  both at LO 
and NLO can be found in \tab{tbl:compMG5xs}. For illustrative purposes
four sample distributions are shown on 
\figs{fig:NLOcomp-pt-tq-a1}{fig:NLOcomp-y-tq-a1}. As it can be seen
from comparing the numbers obtained for cross sections at LO and NLO
and from the distributions the two calculations are in agreement with
each other.

\section{Technical issues}

\subsection{\label{sec:techcuts}Generation cuts}

In a recent work \cite{Kardos:2014zba} we proposed a new method to
generate sufficiently inclusive LHEs including photons, which can be
used to make predictions for the hadroproduction of isolated hard
photons at the hadron level, formally correct at NLO accuracy in
perturbative QCD. The essence of the method is to introduce generation
isolation that is sufficiently loose so that the physical cross section
with usual isolation parameters is independent of the parameters of the
isolation needed for generating the events. For generation isolation
we employ the smooth isolation of Frixione \cite{Frixione:1998jh}
with a small value for the parameter $\delta_0$.  In addition to the
isolation during event generation, we also need a generation cut on the
transverse momentum of each photon, that we set to $5\,\gev$ and
varying it we checked that our physical predictions for hard photon
production do not depend upon this cut.

\subsection{\label{sec:suppress}Suppression factors}

This process has two photons in the final state.  As no photon-photon
splitting is possible two technical cuts are sufficient to obtain a finite
cross section at LO: lower limits  are needed for the transverse
momentum of the two photons. To enhance event generation in the physically
relevant portion of phase space we use a suppression factor which could
be used to suppress events in regions of phase space that are expected
to be cut by physical selection cuts. Instead of this straightforward
generalization of the suppression factor for the \ttgamma\ process
\cite{Kardos:2014zba} we suggest 
\begin{align}
\mathcal{F}_{\rm supp} &= 
\frac{1}{
1 + 
\left(\frac{\pTsupp^2}{\pTgamma{1}^2}\right)^k +
\left(\frac{\pTsupp^2}{\pTgamma{2}^2}\right)^k +
\left(\frac{\msupp^2}{\maa^2}\right)^k
}
\label{eqn:suppfact}
\,,
\end{align}
where \pTgamma{i}\ is the transverse momentum for the $i$th photon,
\maa\ is the invariant mass of the two-photon system and $k=2$ was
chosen throughout. The suppression over the invariant mass of the
two-photon system is introduced to allow for a much better 
population of the large and moderately large \maa\ region.
Our results were obtained with $\pTsupp = 100\,\gev$ and 
$\msupp = 100\,\gev$. 

In the \powhegbox\ the differential cross section correct up to
NLO has the form of
\begin{align}
\ud\sigma^{\rm NLO} &=
\ud\Phi_B B +
\ud\Phi_B \mathcal{V} +
\ud\Phi_B\int\ud\Phi_{\rm rad} (R - C) +
\ud\Phi_B\left(G_\oplus + G_\ominus\right)
\,,
\end{align}
where $B$ is the Born term, $\mathcal{V}$ is the regularized 
virtual contribution, $R$ is the real-emission part, $C$
is a short-hand for all the local subtraction terms regularizing the
real-emission, $G_\oplus$ and $G_\ominus$ are remnants of collinear
factorization, $\ud\Phi_B$ is the underlying Born while $\ud\Phi_{\rm
rad}$ is the one-particle phase space measure and the integral over
momentum fractions is considered implicit. In the \powhegbox\ it is
possible to decompose the real-emission part into two, disjoint
contributions: the singular ($R^{s}$) and remnant ($R^{r}$) ones.
Originally this bisection was introduced to deal with problems arising
when the Born term becomes zero but not all the contributions
\cite{Alioli:2008gx}. This decomposition resulted in faster
event generation and later on became standard \cite{Campbell:2012am}.
With this decomposition the differential cross section takes the form of
\begin{align}
\ud\sigma^{\rm NLO} &=
\ud\Phi_B B +
\ud\Phi_B \mathcal{V} +
\ud\Phi_B\int\ud\Phi_{\rm rad} (R^{s} - C) +
\ud\Phi_B\left(G_\oplus + G_\ominus\right) +
\ud\Phi_R R^{r} =
\nonumber\\
&=
\ud\Phi_B\widetilde{B} + \ud\Phi_R R^{r}
\,,
\end{align}
by introducing $\widetilde{B}$ we can now speak about a $\widetilde{B}$
and a remnant contribution. To decide whether a real-emission contribution
is the singular or the remnant one, the following criterion is used as
default in the \powhegbox:
\begin{equation}
R^s = R - R^r
\,,\qquad
R^r = R\,\Theta\left(R - \xi\max(C_{\mathcal{C}},C_{\mathcal{S}})\right)
\,,
\end{equation}
where $C_{\mathcal{C}}$ is the sum of all the collinear while
$C_{\mathcal{S}}$ is the sum of all the soft subtraction terms and
$\xi$ is an arbitrary parameter set to $5$ as default in the
\powhegbox.  When dealing with a process with photon(s) in the final
state, hence having a generation isolation, large
contributions can turn up when the separation between one of the
photons in the final state and one final state massless quark becomes
smaller than $\delta_0$. These contributions end up
in the remnant provided by the large $R/B$ ratio degrading the
efficiency of remnant-type event generation. This is a consequence of
the method of event generation which is based upon the hit-and-miss
technique: an overall upper bound is determined which is used in a
subsequent step to unweight on the phase space. Large contributions
tend to push this upper bound very high hence large difference can
build up between the upper bound and regular contributions
characterizing the vast majority of phase space hampering the
unweighting procedure.  

When the underlying Born process is already singular close to the
singularity the contributions are enhanced due to the apparent,
unregularized singularity. These large contributions result in a large
norm which forces the hit-and-miss based event generation mechanism to
generate almost all the events close to the singular region. This
problem can be successfully solved by introducing a suppression factor
which, as its name suggests, suppresses these large contributions by
multiplying them with a dynamics-dependent, though small number. In the
generated event file the phase space coverage of the events is such
that less events are generated in and near singular regions. Since
the suppression is a purely non-physical quantity the events can only
bear physical meaning if the event weight incorporates the inverse of
the suppression factor. Thus the suppression of the number of events in
regions bearing no physical importance is compensated by scaling up the
event weight accordingly.  

The concept of the suppression factor can be used to improve the
efficiency of remnant-type event generation. We modify the differential
cross section by introducing a second suppression factor only affecting
the remnant contribution:
\begin{align}
\ud\sigma^{\rm NLO} &=
\ud\Phi_B\widetilde{B} + 
\ud\Phi_R \mathcal{F}_{\rm supp}^{\rm r}(\Phi_R) R^{r}
\,,
\end{align}
where $\mathcal{F}_{\rm supp}^{r}(\Phi_R)$ is the remnant suppression factor
depending upon the real-emission phase space. To obtain physically 
meaningful events when a remnant event is being generated the event weight
has to be multiplied by the inverse of the remnant suppression factor. Since 
large contributions to the remnant part of the cross section only come from
those configurations where a final state (anti)quark is not well-separated from
one of the photons the following form of remnant suppression was used during
our calculations
\begin{align}
\mathcal{F}_{\rm supp}^{r} &=
\prod_{i=1}^{2}\frac{1}{1 + \left(\frac{\pTsuppR}{\pT(\gamma_i,\,q)}\right)^k}
\,,
\end{align}
where $\pTsuppR$ is the remnant suppression parameter set to $20\,\gev$
throughout our calculation, $\pT(\gamma_i,\,q)$ is the relative $\pT$
of the $i$th photon and the final state (anti)quark and we use $k=4$.
The suppression is used only for those contributions where a massless
(anti)quark is present in the final state. The presence and this form of
remnant suppression significantly increased the efficiency of remnant event
generation.

\subsection{\label{subsec:scale}Choice of scale}

\begin{figure}[t]
\centering
\includegraphics[width=0.80\textwidth]{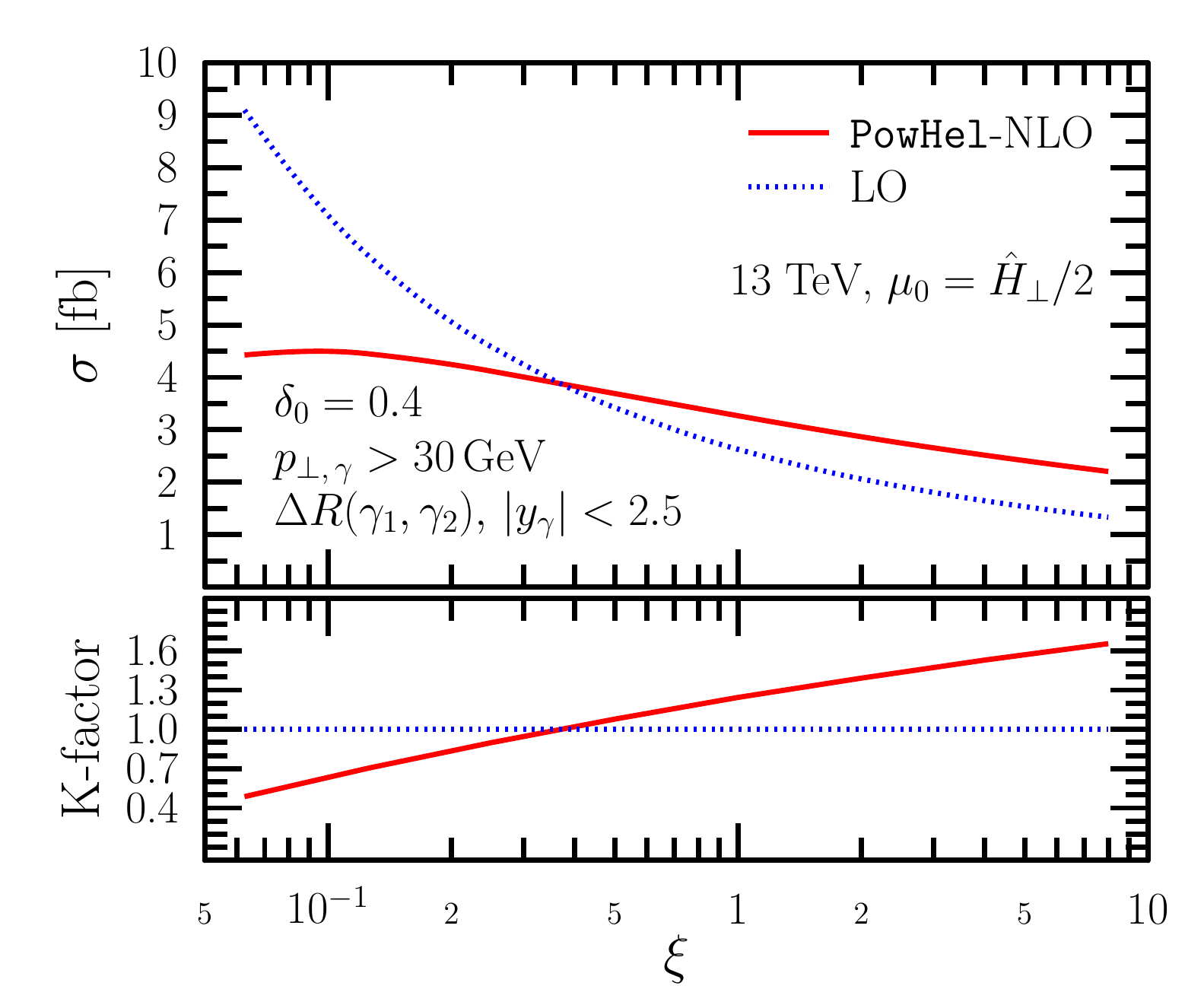}
\caption{{\label{fig:scaledep}} LO and NLO cross sections as functions of
the scale variation parameter. Details of the setup and cuts applied can
be found in the text.}
\end{figure}

The final state for \ttaa\ production contains four particles at the Born
level which leaves us with a rich set of kinematic variables usable for 
setting the renormalization and factorization scales for the process.
In our previous work \cite{Kardos:2013vxa} for the hadroproduction of the
\tT\,\bB\ final state, with massless b-quarks, we found that the half
sum of the transverse masses of final state particles provides a good
scale. It yields a moderate K-factor and small differences in the shapes
of the distributions at LO and NLO accuracies. Thus we adopt the same
choice for \ttaa\ production for the central scale,
\begin{align}
\mu_0 = \hat{H}_\bot / 2 = 
\frac{1}{2}\left(
m_{\bot,\,\rt} +
m_{\bot,\,\bt} +
p_{\bot,\,\gamma_1} +
p_{\bot,\,\gamma_2}
\right)
\,,
\end{align}
where the hat indicates that the underlying Born kinematics was used to
construct the scale. We show the scale dependence of the cross section
as the scale is varied around the central choice in \fig{fig:scaledep}.
To make this prediction we used the following setup: 13~TeV LHC with
\texttt{CT10nlo} as the chosen PDF set using a 2-loop \as, $\mt =
172.5\,\gev$ and the fine-structure constant kept fixed at $\aem =
1/137$. The selection cuts coincide with the set used in \sect{sec:NLO}
with the inclusion of a cut on the separation of the two hard photons
measured in the rapidity azimuthal angle plane, $\Delta
R(\gamma_1,\gamma_2) > 0.4$. The scales are varied by first making the
renormalization and factorization scales coincide, $\mur = \muf = \mu$,
then $\mu$ is varied through $\mu = \xi \mu_0$, where
$\xi\in[1/16,8]$ and $\mu_0 = \hat{H}_\bot/2$. Using this default scale,
the NLO K-factor is $K \simeq 1.24$.
If the scale is varied in the standard region of $[\mu_0/2,\,2\mu_0]$
the scale uncertainty drops from +30\,\%--27\,\% to +14\,\%--13\,\%
if NLO QCD corrections are included. It is interesting to note that
choosing $\mu_0 = \hat{H}_\bot/4$ as default scale, the scale dependence
remains the same, but the K-factor decreases to $K \simeq 1.08$.

\subsection{The effect of the POWHEG Sudakov factor}

\begin{figure}[t]
\includegraphics[width=0.50\textwidth]{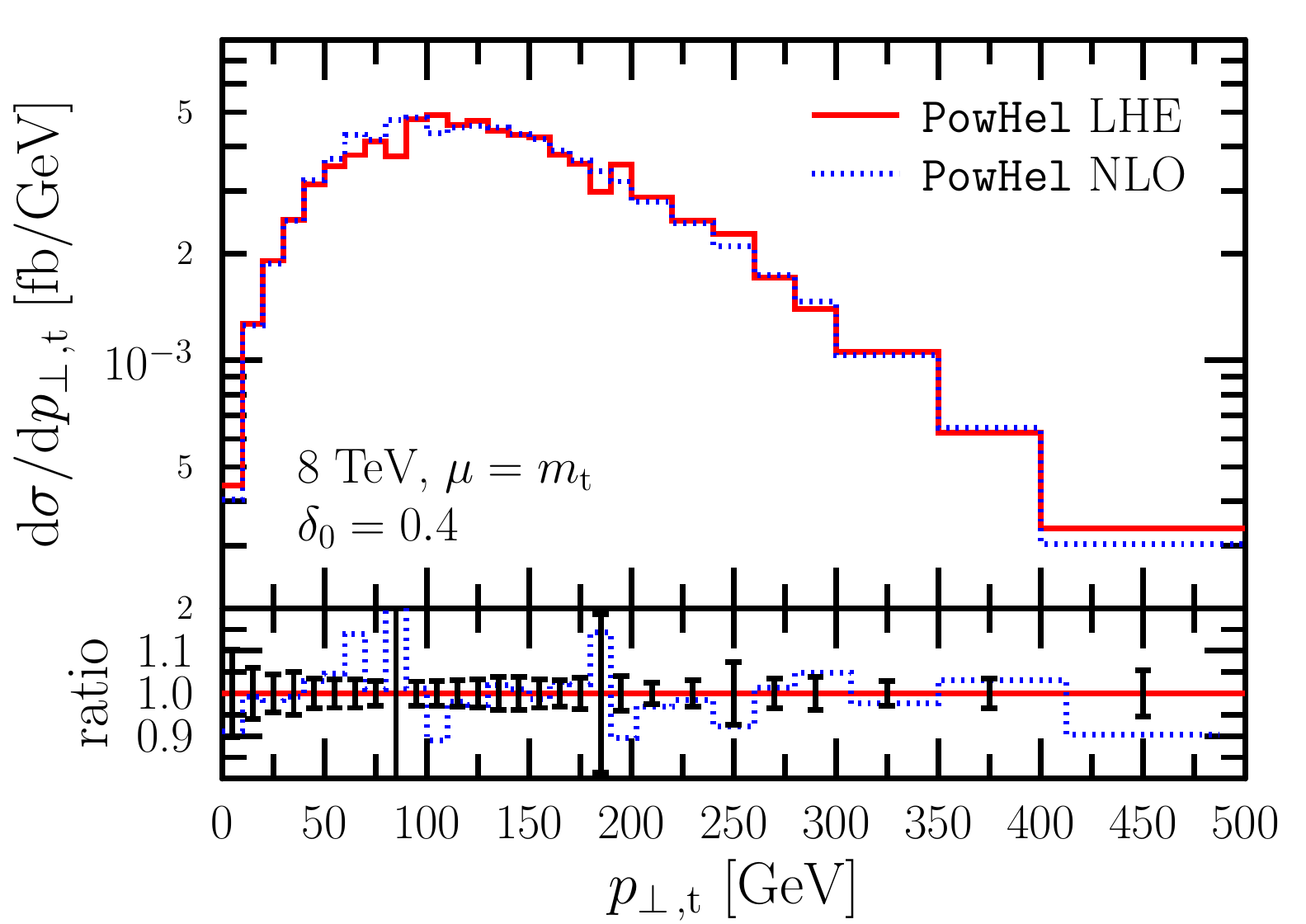}
\includegraphics[width=0.50\textwidth]{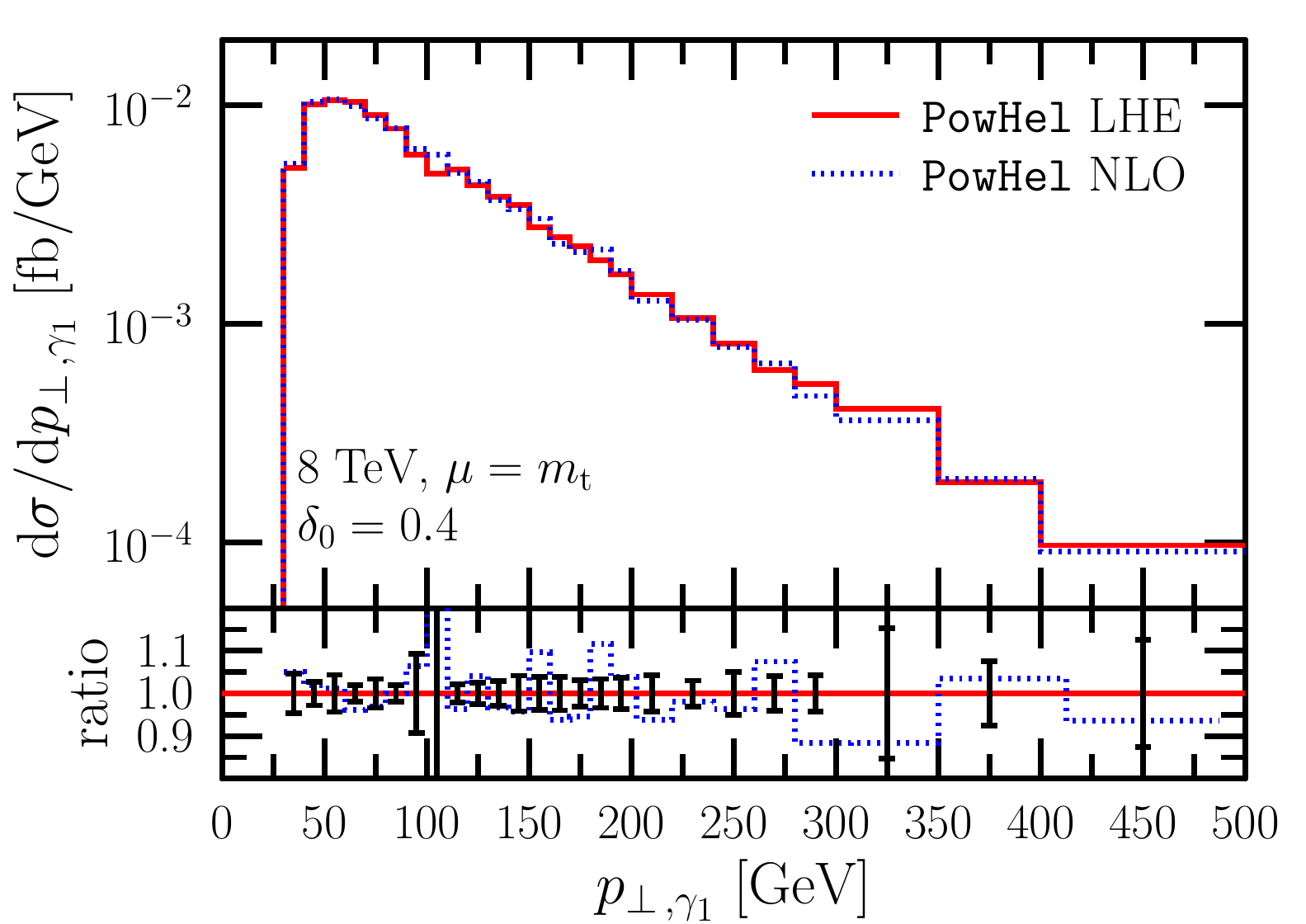}
\caption{\label{fig:NLOLHEcomp-pt-tq-a1} Comparison of the predictions
at NLO accuracy to those obtained from pre-showered events for a
configuration listed in the text for the transverse momentum of the top
quark and for the hardest photon. The lower panel depicts the ratio of
the two predictions.}
\end{figure}

When pre-showered events are generated, the \powheg\ Sudakov factor may
have an effect on measurable quantities through higher order terms in
the perturbative expansion. Hence it is always informative to
compare predictions at the NLO, that is at fixed order, and those
predictions which are obtained from pre-showered events. In order to
quantify the effect of the \powheg\ Sudakov factor events were
generated with the setup of \sect{sec:NLO} and the predictions drawn
from these events were compared to those of the NLO calculation. Sample
distributions can be found on
\figss{fig:NLOLHEcomp-pt-tq-a1}{fig:NLOLHEcomp-ma1a2-logptj}.  From
these distributions, except for the transverse momentum of the extra
parton, a good agreement can be seen between the fixed-order
calculation and predictions drawn from pre-showered events. Taking a
look at the prediction of the transverse momentum of the extra parton
obtained using pre-showered events the usual Sudakov shape can be seen
at smaller values while for large transverse momentum,
$\log_{10}(\pT/\gev) > 1.75$, the LHE prediction merges into the fixed
order one. As we already pointed out in the case of  \ttgamma\
production \cite{Kardos:2014zba} the Sudakov shape gets distorted at
around 0.75 and below due to the presence of the cut (Frixione
isolation) acting on the real-emission part.

\begin{figure}[t]
\includegraphics[width=0.50\textwidth]{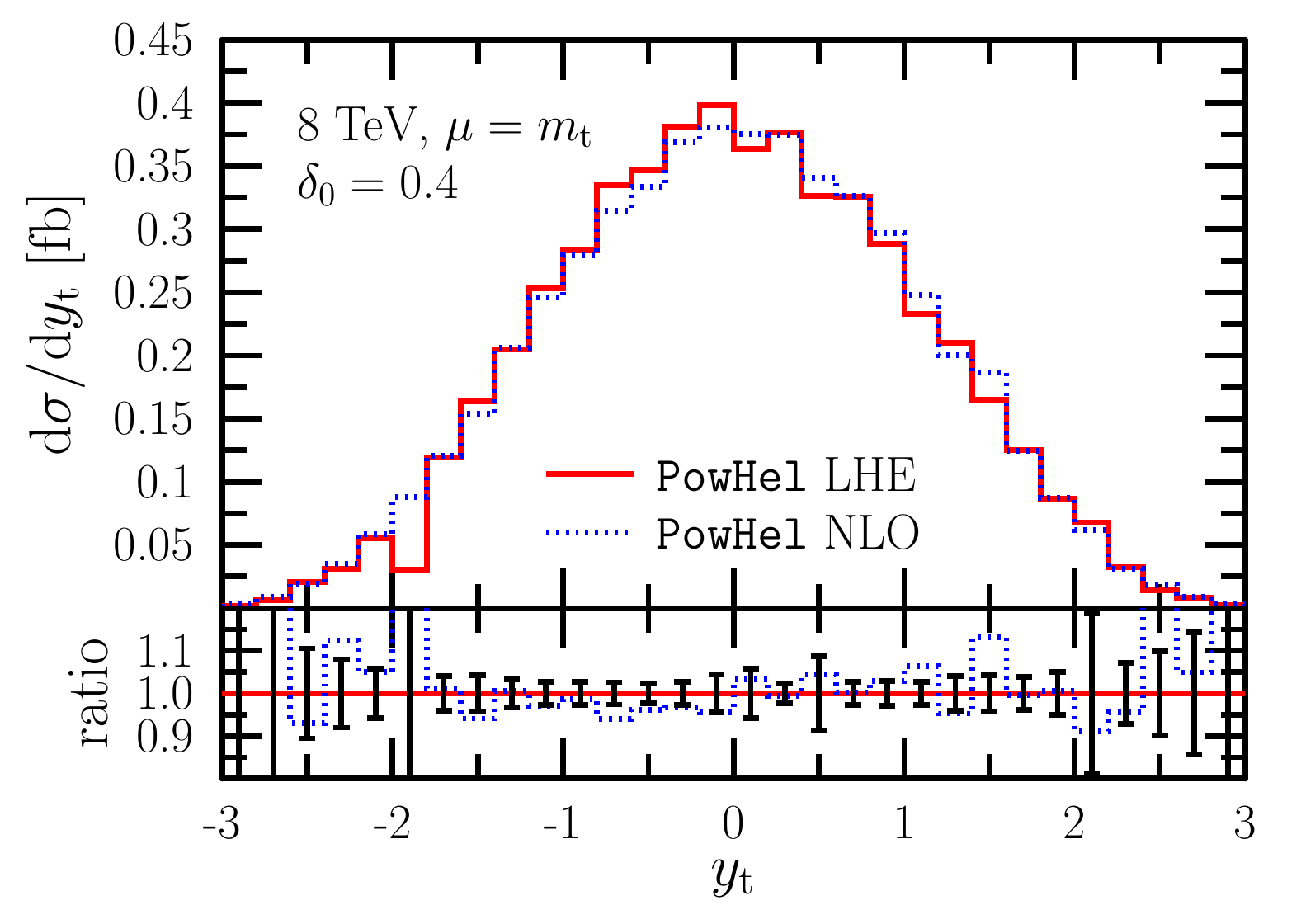}
\includegraphics[width=0.50\textwidth]{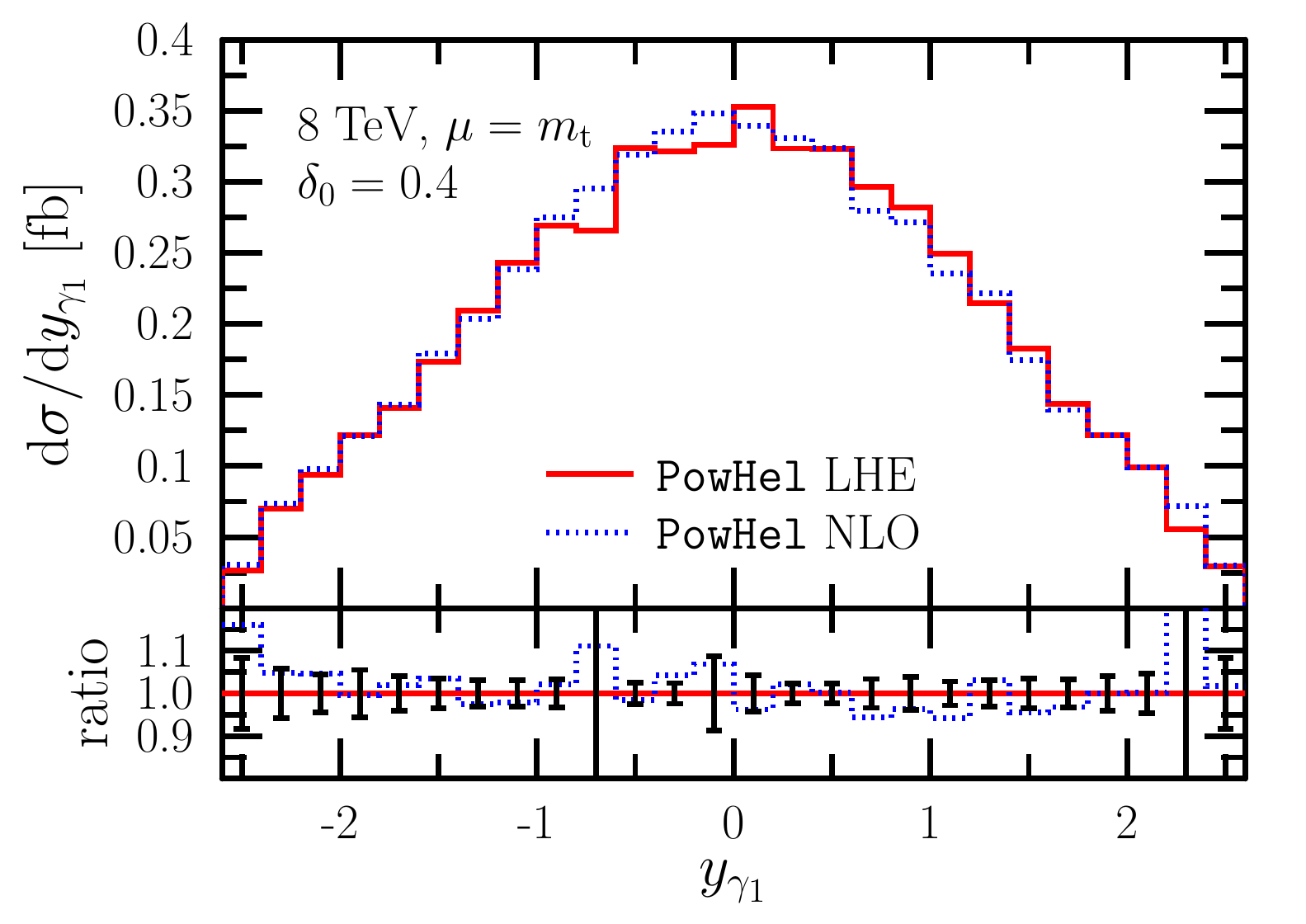}
\caption{\label{fig:NLOLHEcomp-y-tq-a1} The same as 
\fig{fig:NLOLHEcomp-pt-tq-a1} but for the rapidity of the top quark and the
hardest photon.}
\end{figure}

\begin{figure}[t]
\includegraphics[width=0.50\textwidth]{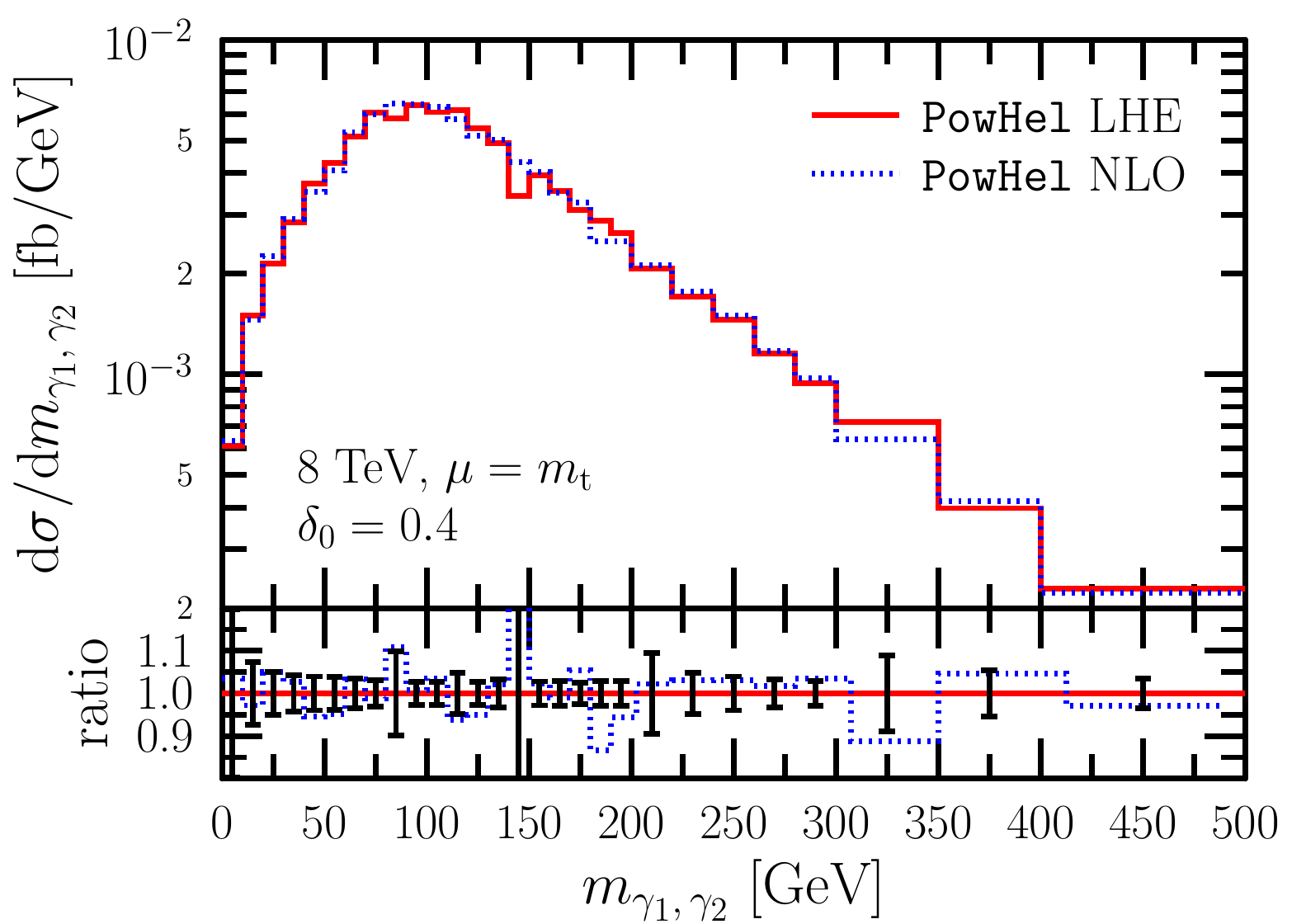}
\includegraphics[width=0.50\textwidth]{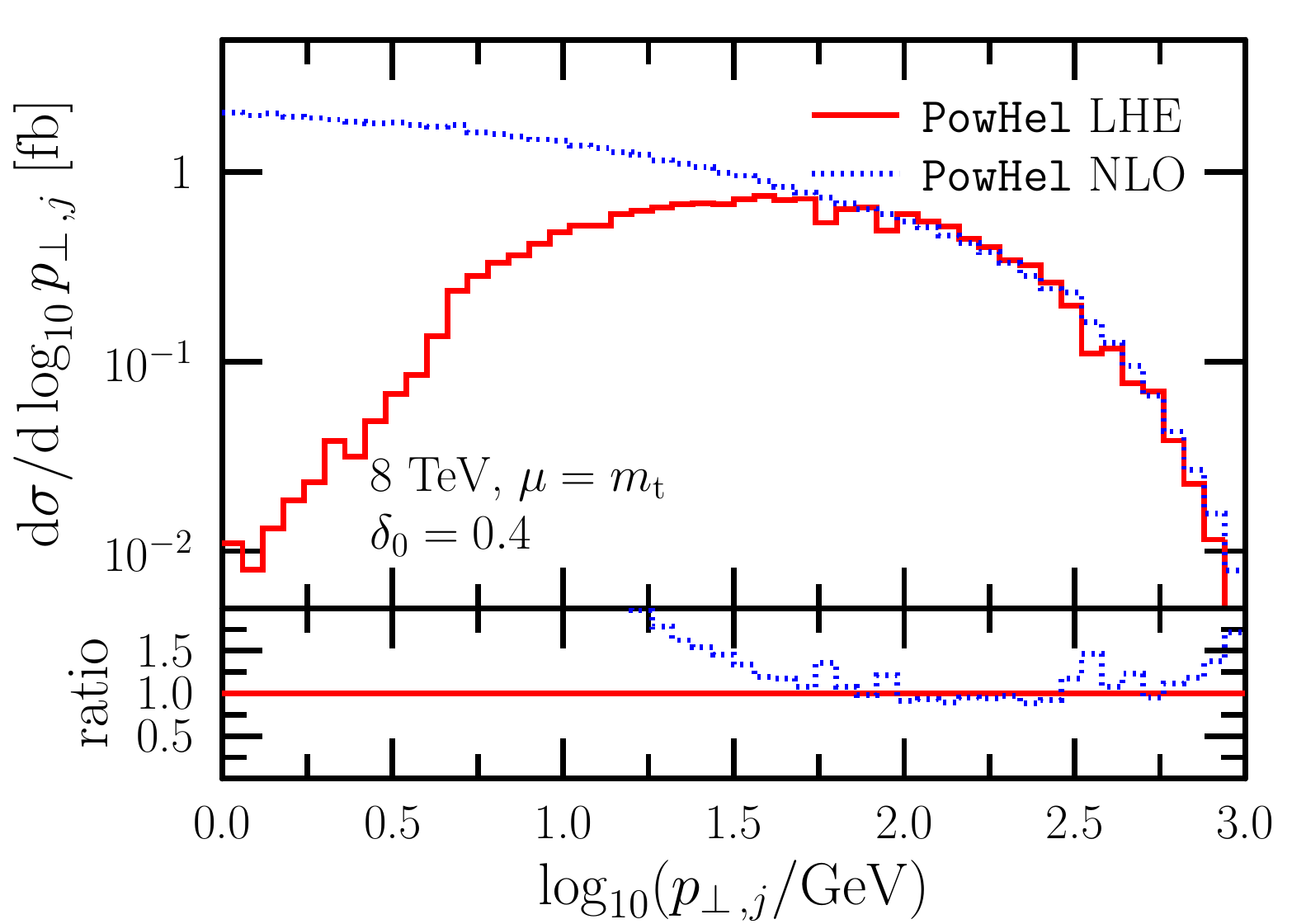}
\caption{\label{fig:NLOLHEcomp-ma1a2-logptj} The same as 
\fig{fig:NLOLHEcomp-pt-tq-a1} but for the invariant mass of the diphoton
system and the transverse momentum of the extra parton.}
\end{figure}

\section{\label{sec:TechIsolIndep} Independence of the generation isolation}

\begin{table}[t]
\begin{center}
\begin{tabular}{|c|c|c|c|}
\hline
\hline
$\delta_0$ & $\sigma^{\rm LHE}$ [fb] & $\sigma^{\rm PS}$ [fb] &
$\sigma^{\rm SMC}$ [fb] \bigstrut\\
\hline
\hline
$0.01$ & $3.31\pm 0.03$ & $3.03\pm 0.04$ & $2.17\pm 0.02$ \bigstrut\\
\hline
$0.05$ & $3.30 \pm 0.03$ & $3.07 \pm 0.05$ & $2.13 \pm 0.02$ \bigstrut\\
\hline
$0.1$  & $3.33 \pm 0.03$ & $3.05 \pm 0.03$ & $2.18 \pm 0.03$ \bigstrut\\
\hline
\hline
\end{tabular}
\end{center}
\caption{{\label{tbl:xscomp}}Cross sections obtained at different stages
of event evolution for three different values of generation isolation
using the cuts listed in the text.}
\end{table}

\begin{figure}[t]
\includegraphics[width=0.50\textwidth]{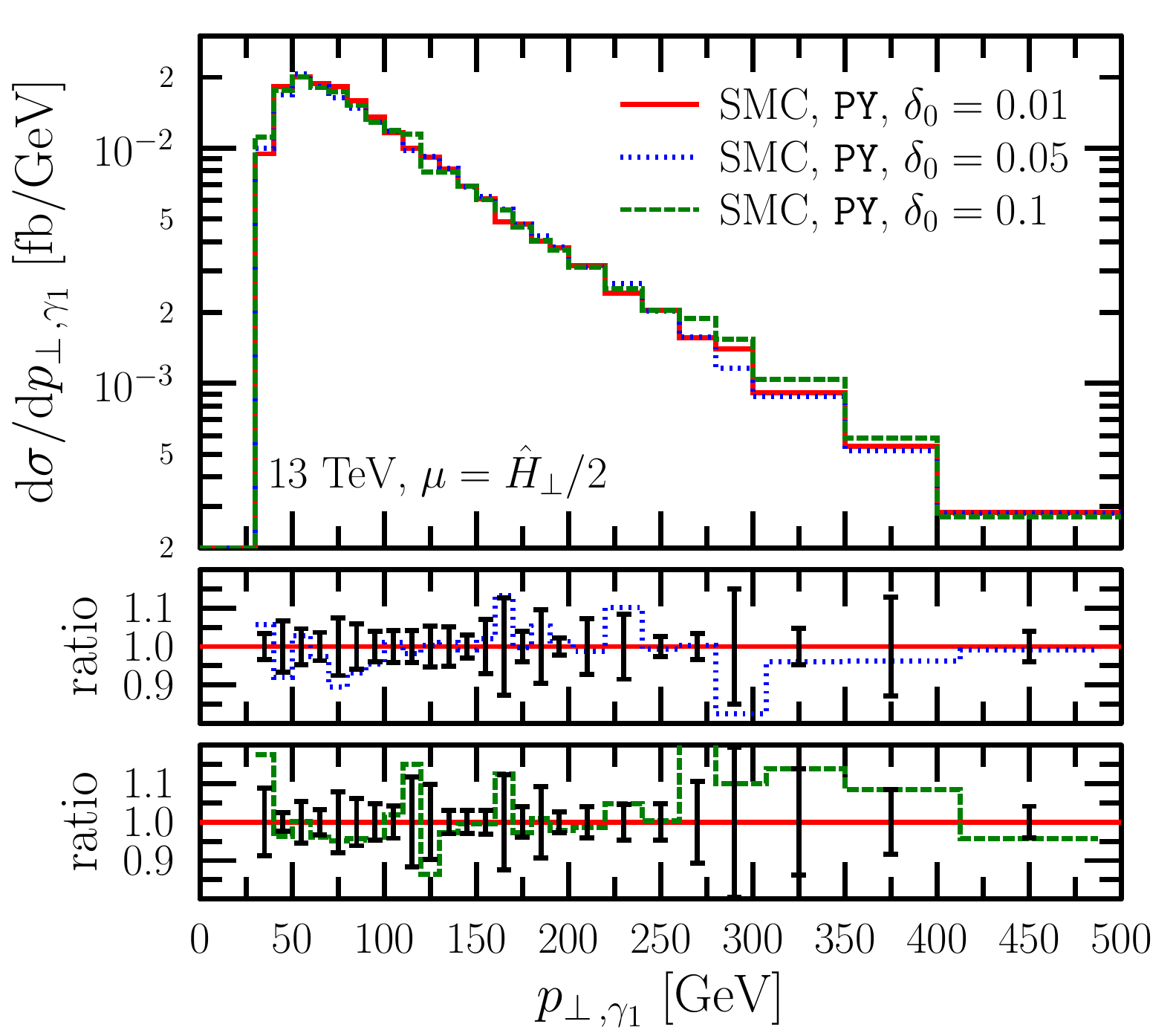}
\includegraphics[width=0.50\textwidth]{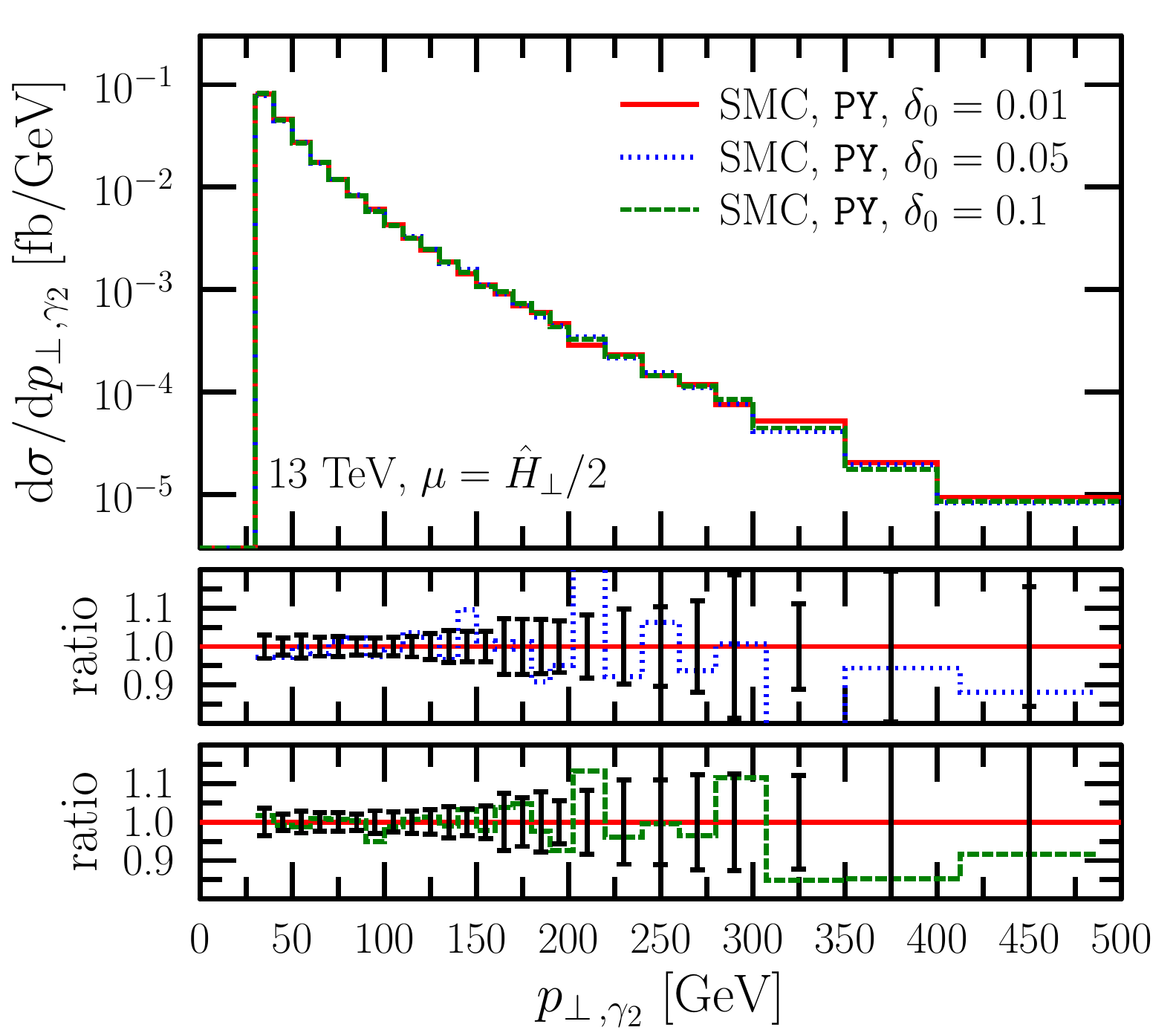}
\caption{\label{fig:SMCtechcomp-pt-a1-a2} Comparison among predictions at the
hadron level taken with three different generation isolation, 
$\delta_0 = 0.1$ (green dashed), 0.05 (blue dotted) and 0.01
(red solid), for the transverse momenta of the two hardest photons.
The lower panels show the ratios with respect to the case of $\delta_0 = 0.01$.}
\end{figure}

In our previous work \cite{Kardos:2014zba} we carried out an extensive
study to prove the independence of generation isolation at various
stages of evolution of the pre-showered events. For the present process
we have repeated all steps, but we restrict ourselves to present the
case of full SMC. At other stages of event evolution the predictions
taken with different generation isolations are in agreement with each
other. The events were generated for the setup listed in
\sect{subsec:scale}. For concreteness the cross sections obtained at
various levels for different generation isolations are listed in
\tab{tbl:xscomp}. These cross sections correspond to the following set
of cuts:
\begin{itemize}
\itemsep=-1pt
\item Jets are reconstructed with the IR-safe anti-\kT\ algorithm
\cite{Cacciari:2008gp} with $R = 0.4$ and $\pTj > 30\,\gev$ using \fastjet 
\cite{Cacciari:2011ma,Cacciari:2005hq}.
\item Two hard photons, $\pTgamma{} > 30\,\gev$, are requested in 
the central region, $|y_\gamma| < 2.5$.
\item The photons should be isolated from the jets and each other such
that $\Delta R > 0.4$. The separation is measured in the rapidity-azimuthal
angle plane.
\item The hadronic activity is limited around both photons in a cone of
$R_\gamma = 0.4$: $E_{\bot,\,{\rm had}}^{\rm max} = 3\,\gev$.
\end{itemize}
The hadronic activity in a cone of $R_\gamma$ around the photon
momentum, $p_\gamma$, is measured through the total hadronic
transverse energy deposited in this cone:
\begin{align}
E_{\bot,\,{\rm had}} =
\sum_{i\in{\rm tracks}} E_{\bot,\,i}
\Theta\left(R_\gamma - R(p_\gamma,p_i)\right) < E_{\bot,\,{\rm had}}^{\rm max}
\,,
\end{align}
where the summation runs over all the hadronic tracks. To make our
predictions at the hadron level we used \pythiaver.  At the hadron
level we allowed tops to decay, but kept QED shower and multiparticle
interactions inactivated. To see the independence of generation
isolation we generated events with three different smooth isolation
parameters ($\delta_0 \in \{0.1,\,0.05,\,0.01\}$) and compared
these pairwise. Sample distributions at the hadron level are depicted
on \figss{fig:SMCtechcomp-pt-a1-a2}{fig:SMCtechcomp-y-a1-tq}.  Among
these distributions t-quark related observables can be found too. These
were obtained by reconstructing the t-quark using \texttt{MCTRUTH}.
From these distributions we can see that the deviation of predictions
from each other is only due to statistical fluctuations. All of the
event sets taken with the three different generation isolations
result in the same predictions at the hadron level which means that
any of these can be safely used in analyses to provide theoretical
predictions.  

\begin{figure}[!ht]
\includegraphics[width=0.50\textwidth]{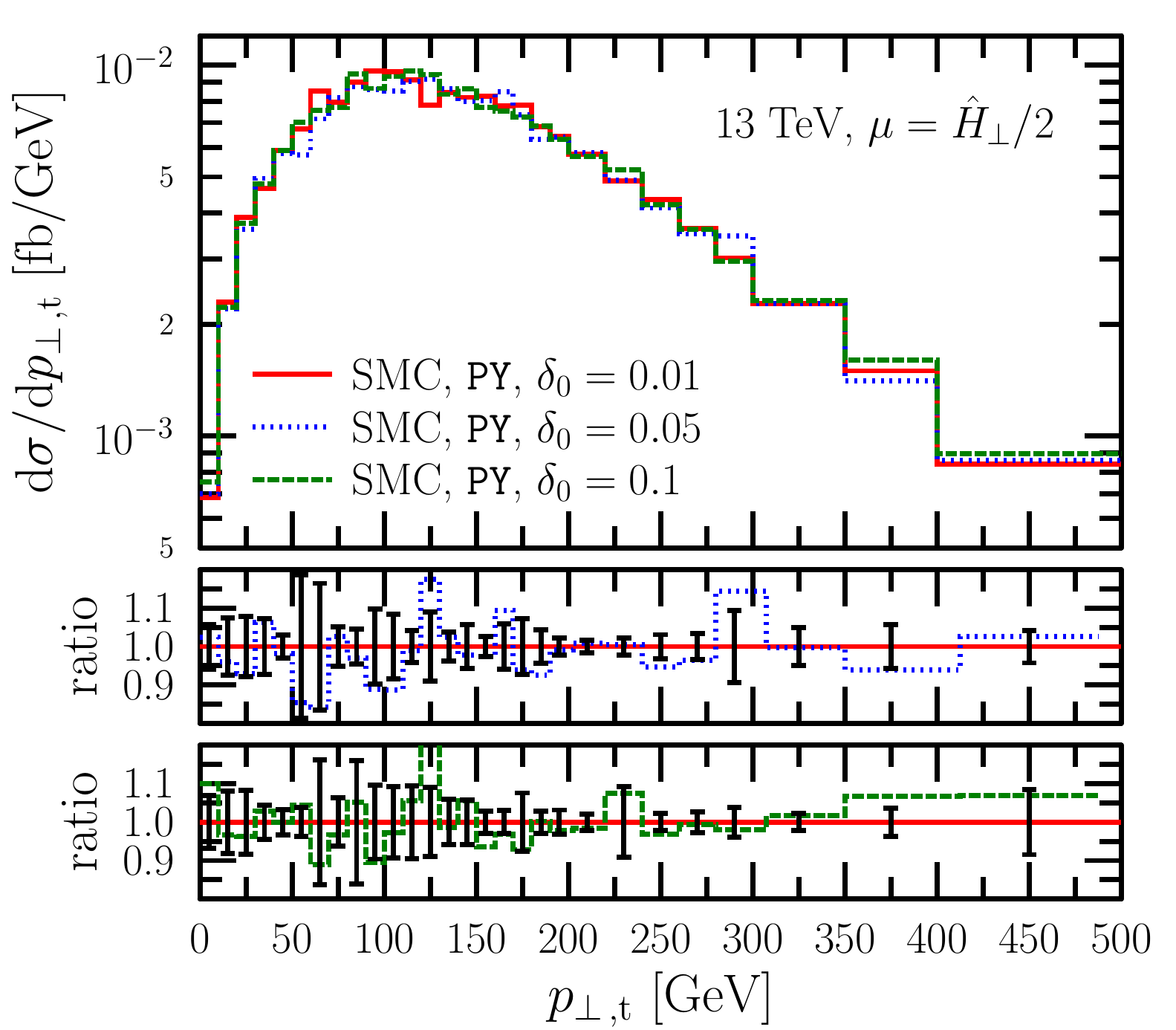}
\includegraphics[width=0.50\textwidth]{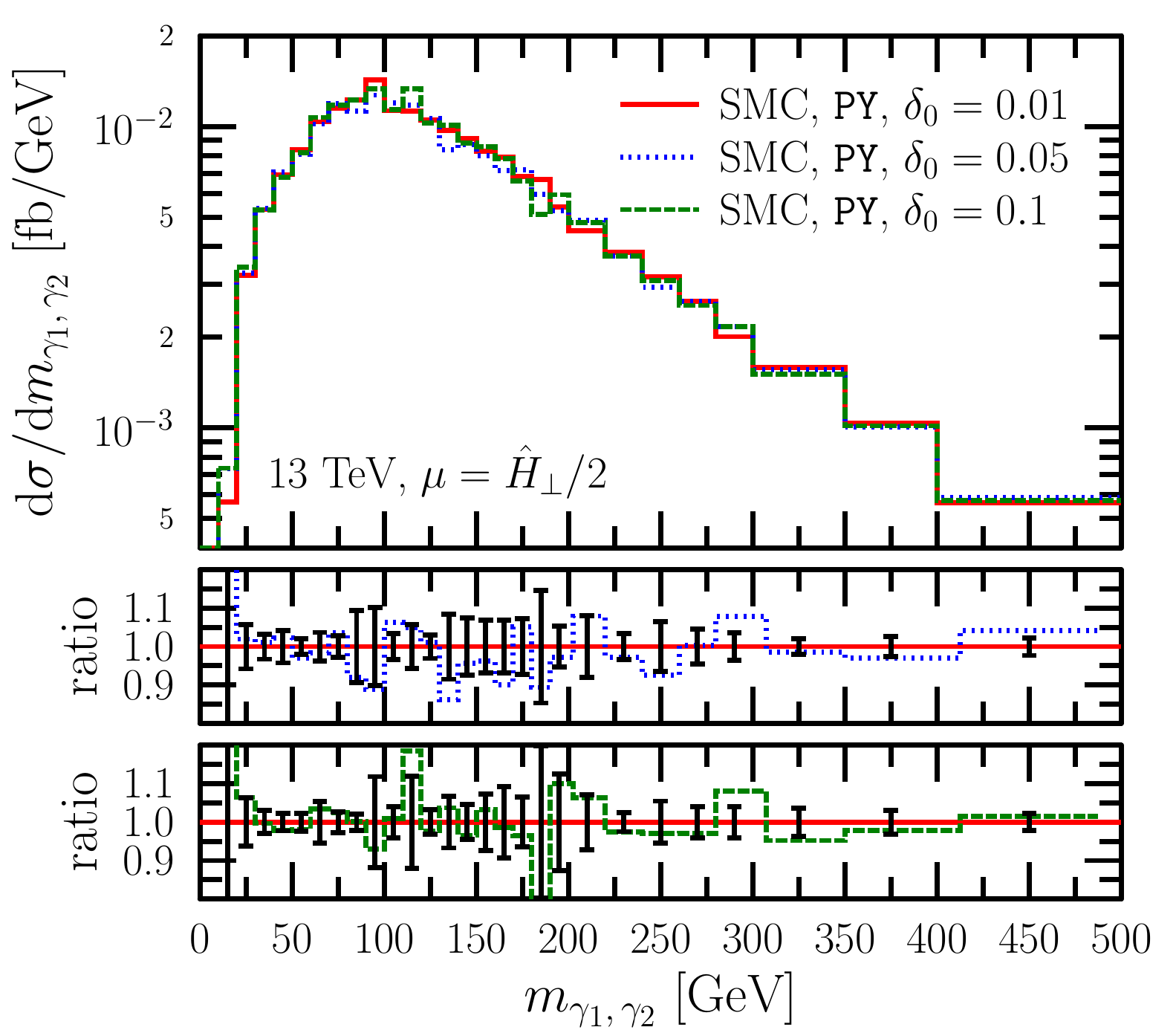}
\caption{\label{fig:SMCtechcomp-pt-tq-m-a1a2} The same as 
\fig{fig:SMCtechcomp-pt-a1-a2} but for the transverse momentum of the top 
quark and the two-photon invariant mass.}
\end{figure}

\begin{figure}[!ht]
\includegraphics[width=0.50\textwidth]{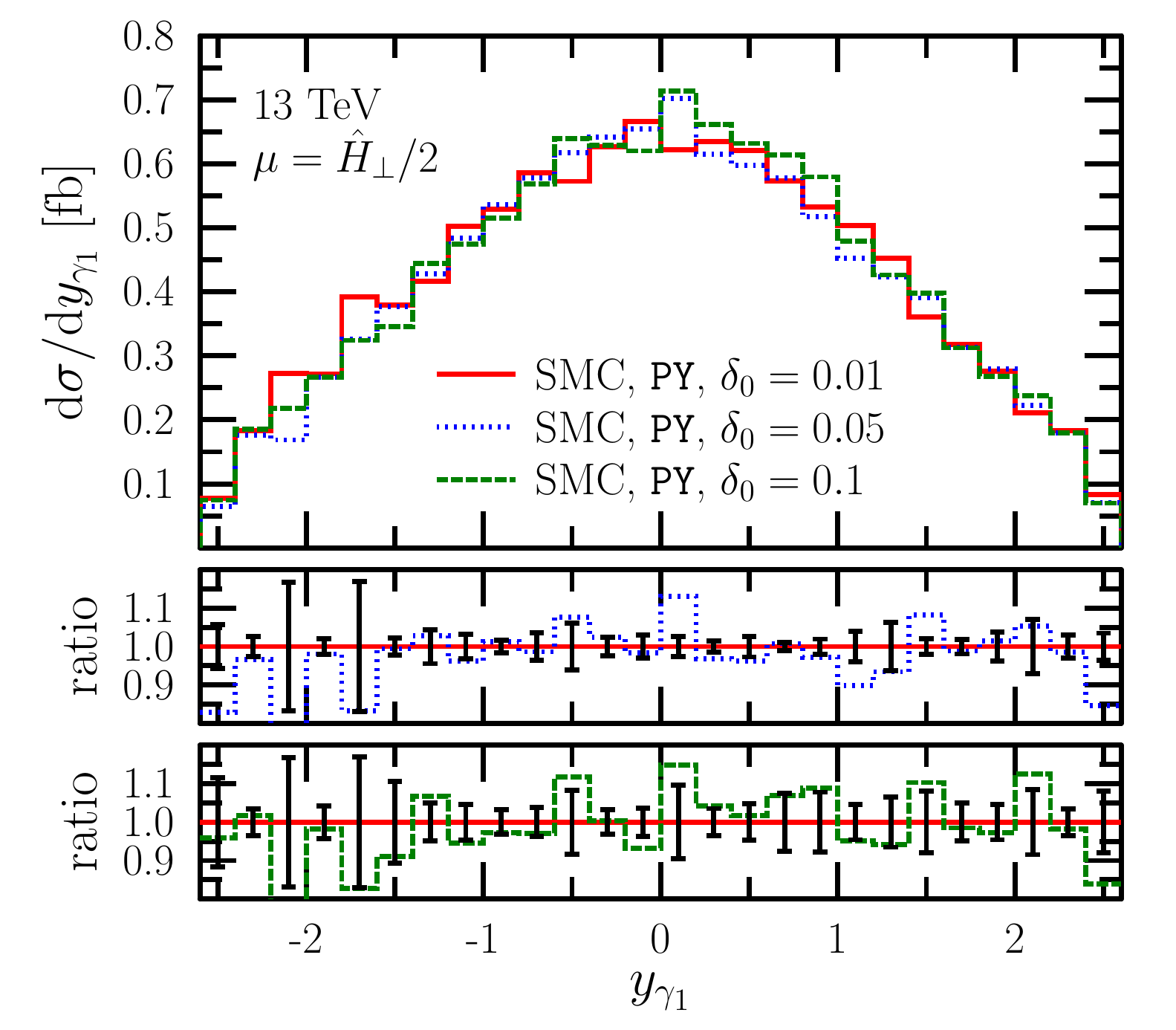}
\includegraphics[width=0.50\textwidth]{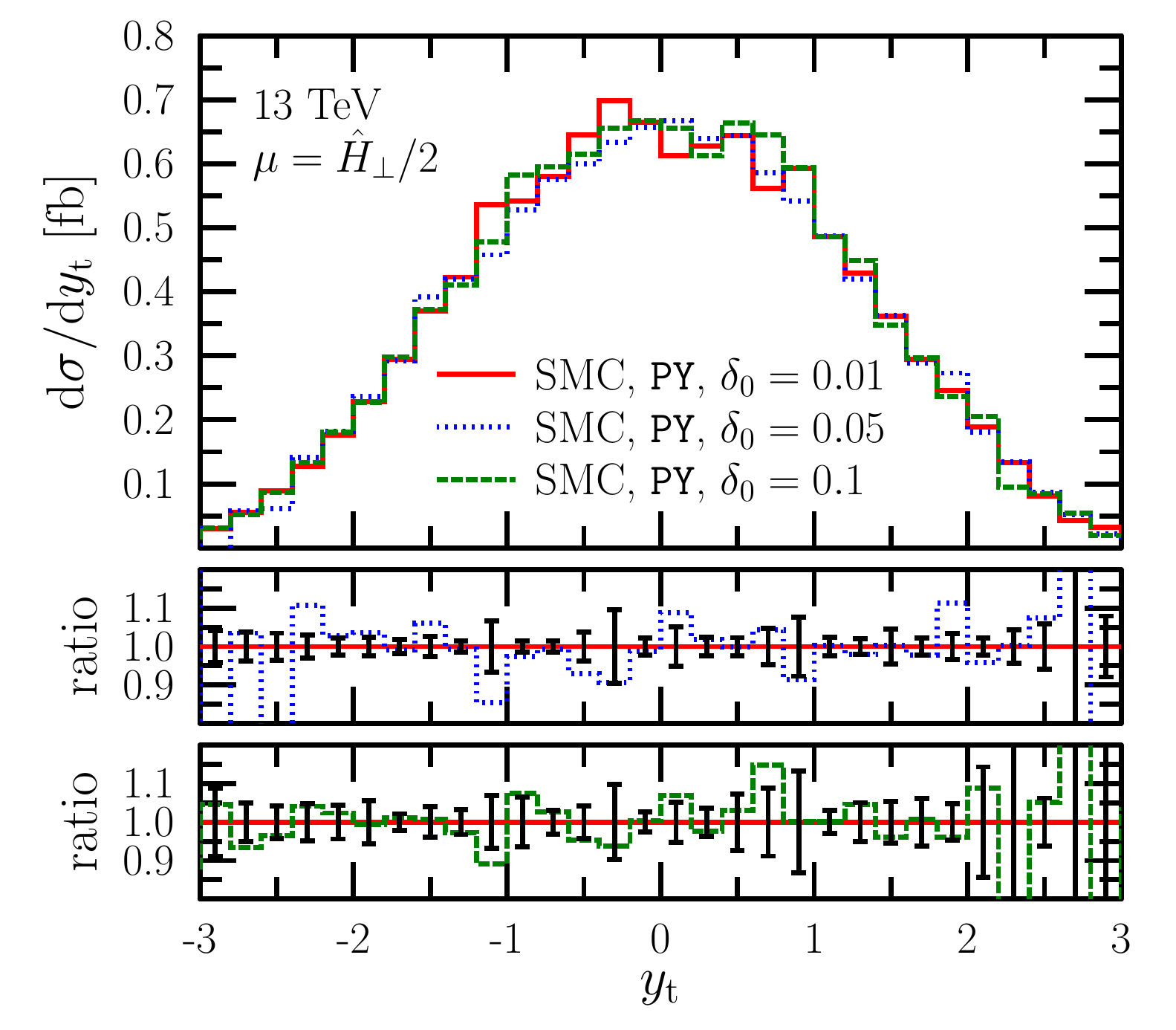}
\caption{\label{fig:SMCtechcomp-y-a1-tq} The same as 
\fig{fig:SMCtechcomp-pt-a1-a2} but for the rapidity of the hardest photon
and the top quark.}
\end{figure}

\section{\label{sec:pheno} Phenomenology}

The main consequence of the previous two sections is the ability to
make physical predictions with experimental isolation criteria using
pre-showered events prepared with a generation isolation chosen suitably
small. The \ttaa\ production plays an important role in Higgs physics
when attention is turned to the properties of the Higgs boson. One such
important property is the Yukawa coupling of the Higgs  boson to various
massive fermions. The Yukawa coupling of the Higgs boson to the top quark can
be measured in \tth\ production. When the Higgs boson decays in the diphoton
channel the irreducible background is \ttaa\ production.  As we
provided predictions for \tth\ production in Ref.\
\cite{Garzelli:2011vp} at the hadron level, we decided to present
predictions for both the signal and the background at the hadron level
with the highest available precision.  For the \ttaa\ background we
take the events generated with $\delta_0 = 0.05$. For the signal we generated a
new bunch of pre-showered events sharing the same parameters with the
\ttaa\ sample, generated for the 13~TeV LHC with \texttt{CT10nlo} PDF
and accordingly chosen 2-loop \as, $\mt=172.5\,\gev$, $\mh=125\,\gev$.
We chose the renormalization and factorization scales equal to each
other and set to $\mur = \muf = \mt + \mh/2$. To make our predictions,
we used \pythiaver\ for simulating the evolution of the events to the
hadron level, but with multiple interactions turned off.

\begin{figure}[t]
\includegraphics[width=0.50\textwidth]{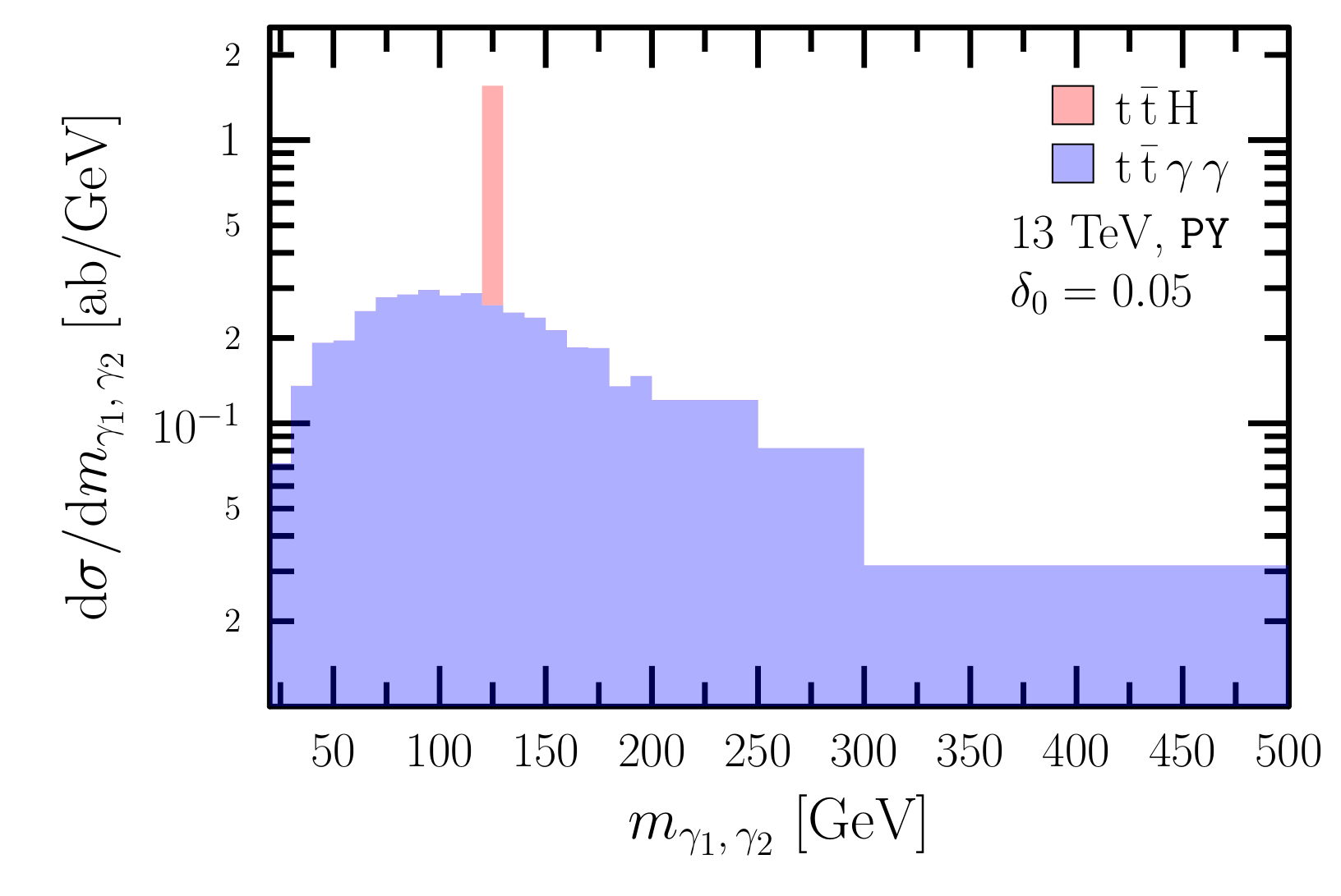}
\includegraphics[width=0.50\textwidth]{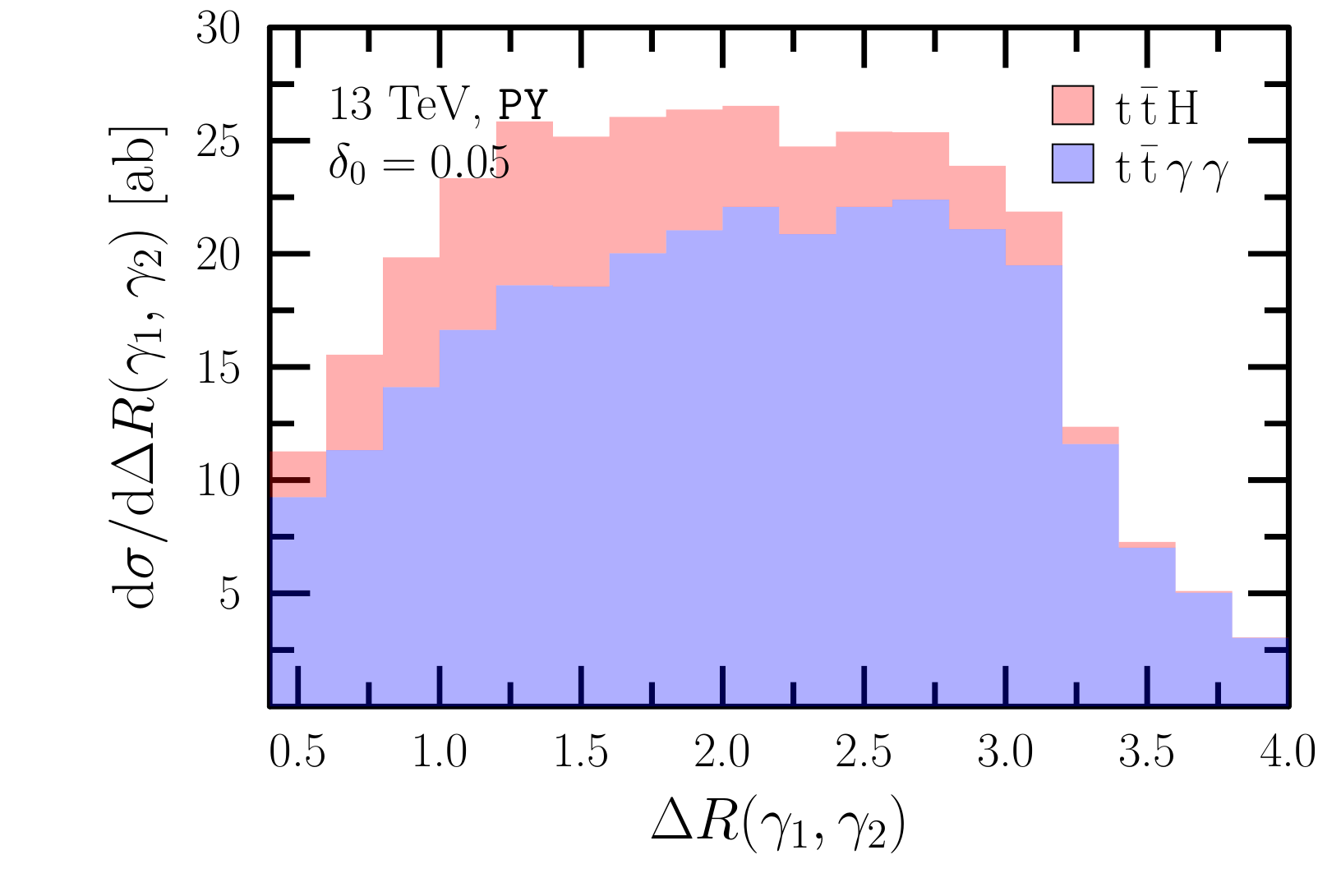}
\caption{\label{fig:SMCpheno-m-dR-a1-a2} The invariant mass and rapidity
azimuthal separation is shown for the diphoton system at the hadron level
for the signal (\tth, red) and for the background (\ttaa, blue) such that 
the signal sits on the top of the background. Further details of the 
calculation can be found in the main text.}
\end{figure}

The partonic final state contains a t-quark pair for both the signal
and the background. There are two ways to detect t-quarks: either
through their decay products, by looking for displaced secondary
vertices as tagging for decaying b-quarks, or using t-tagging. We
decided to use the latter as provided by the \heptoptagger\
\cite{Plehn:2010st,Plehn:2011sj}. In order to perform t-tagging we
followed the following steps:
\begin{itemize}
\item Jet reconstruction using all the hadronic tracks with
the C/A algorithm with $R = 1.5$ using \fastjet
\cite{Cacciari:2011ma,Cacciari:2005hq}. 

\item Only those jets were considered for which 
$\pT > 200\,\gev$ and $|y| < 5$.

\item The t-quark candidate subjets were looked for in the jet mass
range of 150~--~200\,GeV.

\item We selected that particular subjet as a t-quark jet for which
the jet mass was the closest to \mt.
\end{itemize}
In respect to other parameters concerning the top tagging we kept
the default values provided by the \heptoptagger. The cuts applied
to the hadronic events were the following:
\begin{itemize}
\item Two hard photons were requested in the central region with
$\pTgamma{} > 30\,\gev$ and $|y_\gamma| < 2.5$.

\item The photons should be isolated from each other: 
$\Delta R(\gamma_1,\,\gamma_2) > 0.4$.

\item To isolate the photons from hadronic activity a jet 
clustering was performed on all the hadronic tracks with the
anti-\kT\ algorithm \cite{Cacciari:2008gp} as implemented in 
\fastjet \cite{Cacciari:2011ma,Cacciari:2005hq} with
$R=0.4$ and $\pTj > 30\,\gev$. The photons were requested to
be isolated from these jets by $\Delta R(\gamma,j) > 0.4$ measured on the
rapidity-azimuthal angle plane.

\item Both of the hard photons had to be isolated from the top
jet obtained by top tagging and from the three subjets of the
top jet by $\Delta R(\gamma,j) > 0.4$ measured on the rapidity-azimuthal
angle plane.

\item One hard lepton was requested in the central region with 
$\pTlepton{} > 30\,\gev$ and $|y_\ell| < 2.5$. To select leptons we did not
make any distinction between leptons and antileptons.

\item The lepton had to be isolated from both the jets and the photons
with $\Delta R(\ell,i) > 0.4,\,i\in\{\gamma,j\}$ measured on the 
rapidity-azimuthal angle plane.

\item Around both photons a minimal hadronic activity was allowed in a 
cone of $R = 0.4$ such that $E_{\perp,\,{\rm had}}^{\rm max} = 3\,\gev$.
\end{itemize}
In order to reach reasonable statistics even for the signal we turned off
all decay channels of the Higgs boson but the diphoton one and rescaled
the event weights with the $H\to\gamma\,\gamma$ branching ratio. The
most interesting distribution is the invariant mass of the two-photon
system along with this a couple more distributions are depicted in 
\figs{fig:SMCpheno-m-dR-a1-a2}{fig:SMCpheno-pt-a1-a2}. By taking a look
at the distributions it is clear that the cross section drops into the
attobarn range.  Due to the narrow width of the Higgs boson the signal
appears as a single, well-defined spike in the $m_{\gamma\gamma}$
spectrum with excellent signal/background ratio. For all the other
distributions the background overwhelms the signal.  

\begin{figure}[t]
\includegraphics[width=0.50\textwidth]{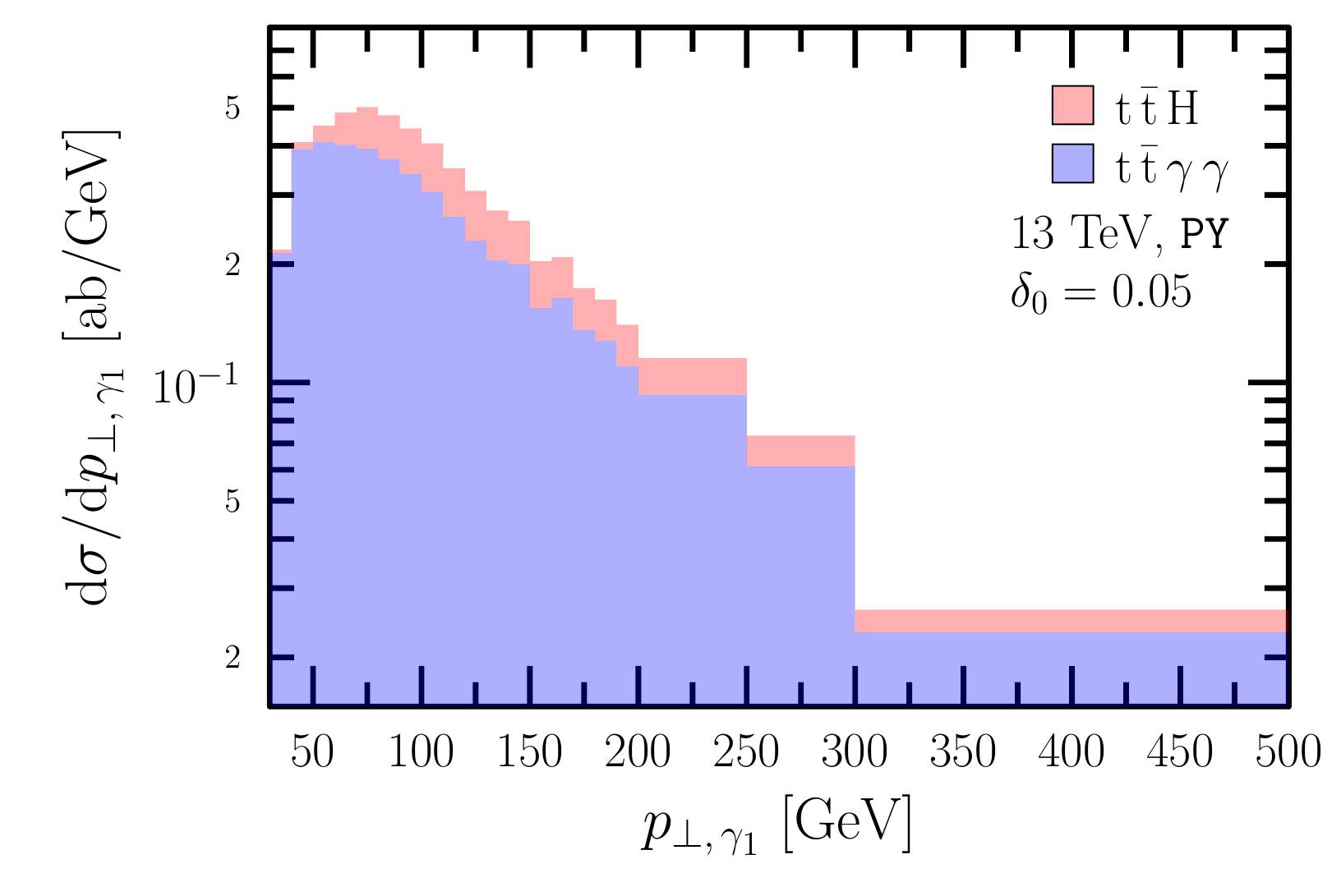}
\includegraphics[width=0.50\textwidth]{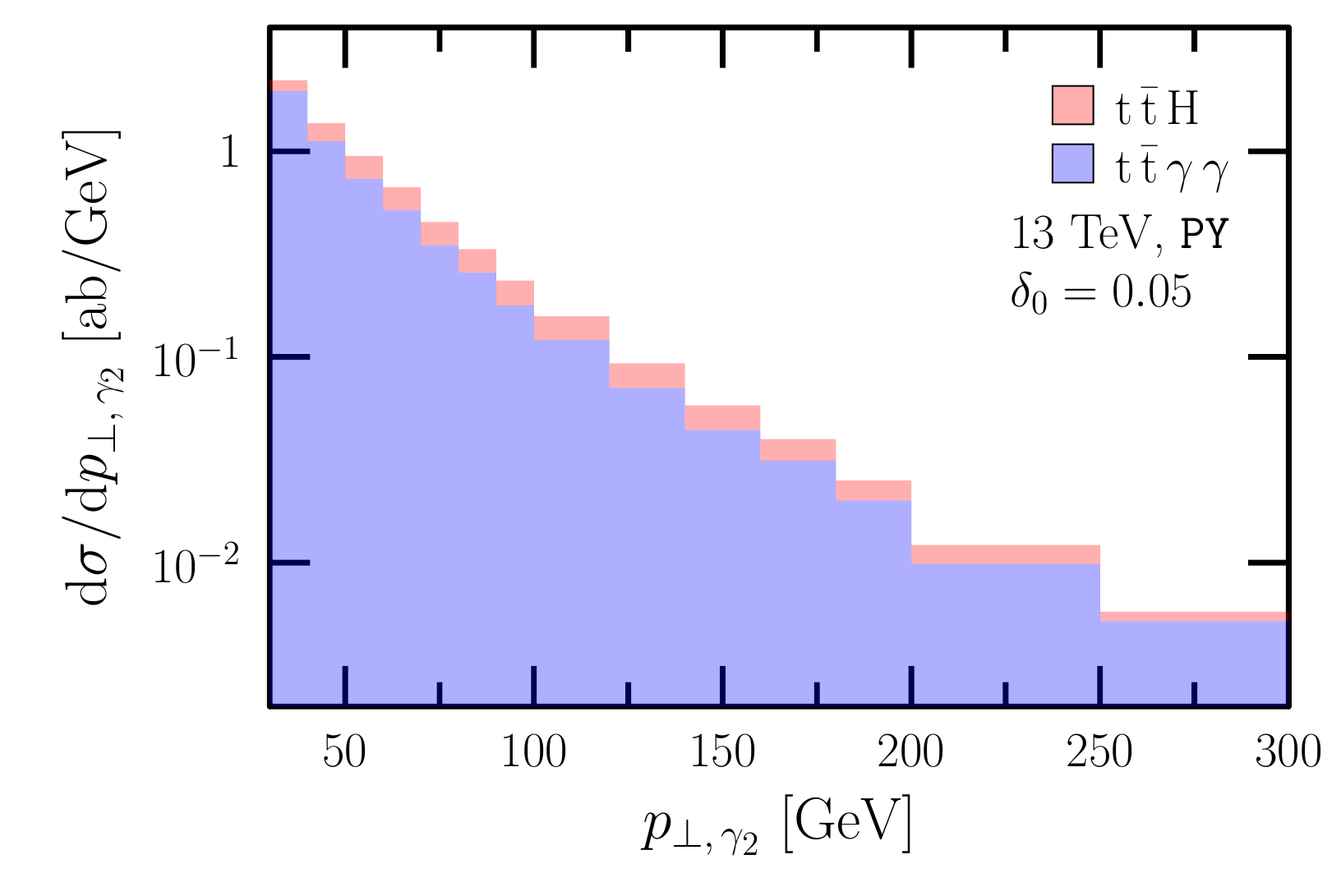}
\caption{\label{fig:SMCpheno-pt-a1-a2} The same as \fig{fig:SMCpheno-m-dR-a1-a2}
but for the transverse momentum of the two hard photons.}
\end{figure}

\section{\label{sec:conclusions} Conclusions}

We presented the first predictions for the hadroproduction of \ttaa\
final state both at NLO accuracy and at NLO matched with parton shower.
The predictions at NLO were computed by two programs, \powhel\ and
\madgraphfive\ and complete agreement was found. While the predicitions
at NLO accuracy can be made only using smooth isolation prescription of
Frixione for the hard photons, the matched predictions are based on
LHEs that can be exposed to the usual cone-type isolation of
experimenters. This is achieved by generating the events using
generation cuts that are sufficiently loose, so that the predictions do
not depend on those when usual physical isolations are used.

We found that using half of the sum of the transverse masses of
particles in the final state the NLO corrections are about 25\,\%, and
decrease the dependence on the renormalization and factorization scales
significantly. The generated events are stored and can be downloaded from
http://www.grid.kfki.hu/twiki/bin/view/DbTheory/TtaaProd. We generated
similar events for the \ttH\ final state, to which the \ttaa\ process is
an irreducible background in the $H\to \gamma\gamma$ decay channel. We
compared some kinematic distributions of both signal and background
process and found that the invariant mass of the two hard photons shows
a clearly visible narrow peak, but with small cross section, in the
attobarn range for the LHC at 13\,TeV. For other kinematic
distributions the signal and background have rather similar shapes.

\noindent{\bf Acknowledgements}
This research was supported by
the Hungarian Scientific Research Fund grant K-101482,
the SNF-SCOPES-JRP-2014 grant
``Preparation for and exploitation of the CMS data taking at the next LHC run'',
the European Union and the European Social Fund through
LHCPhenoNet network PITN-GA-2010-264564,
and the
Supercomputer, the national virtual lab TAMOP-4.2.2.C-11/1/KONV-2012-0010
project.

\providecommand{\href}[2]{#2}\begingroup\raggedright\endgroup

\end{document}